\documentclass[12pt]{article}
\pdfoutput=1

\usepackage{putex}
\usepackage{comment}
\usepackage{graphicx}
\usepackage{caption}
\usepackage{amsmath}
\usepackage{array}
\usepackage{subcaption}
\usepackage{epstopdf}
\usepackage{enumerate}
\usepackage{cite}
\usepackage{youngtab}
\usepackage{tensor}
\usepackage{slashed}
\usepackage[aligntableaux=center]{ytableau}
\usepackage[utf8]{inputenc}
\usepackage[
      colorlinks=true,
      linkcolor=blue,
      urlcolor=blue,
      filecolor=black,
      citecolor=red,
      ]{hyperref}

\newcommand{\HH}{\mathbb{H}}

\newcommand{\abs}[1]{\left\lvert #1 \right\rvert}

\newcommand {\be} {\begin {equation}}
\newcommand {\ee} {\end {equation}}

\newcommand {\bes} {\begin {equation*}}
\newcommand {\ees} {\end {equation*}}

\newcommand{\es}[2] {\begin{equation} \label{#1} \begin{split} #2 \end{split} \end{equation}}

\newcommand{\R}{\mathbb{R}}

\newcommand{\beq}{\begin{equation}}
\newcommand{\eeq}{\end{equation}}

\def\ie{\begin{equation}\begin{aligned}}
\def\fe{\end{aligned}\end{equation}}

\numberwithin{equation}{section}

\def\<{\langle}
\def\>{\rangle}
\def \eps {\epsilon}

\begin{document}

\preprint{PUPT-2599}

\institution{PU}{Joseph Henry Laboratories, Princeton University, Princeton, NJ 08544, USA}
\institution{PCTS}{Princeton Center for Theoretical Science, Princeton University, Princeton, NJ 08544}
\institution{HU}{Department of Physics, Harvard University, Cambridge, MA 02138}

\title{The $O(N)$ Model in $4<d<6$:\\ Instantons and Complex CFTs}

\authors{Simone Giombi,\worksat{\PU} Richard Huang,\worksat{\PU} Igor R.~Klebanov,\worksat{\PU,\PCTS} Silviu S.~Pufu,\worksat{\PU} \\[10pt] and Grigory Tarnopolsky\worksat{\HU}}

\abstract{
We revisit the scalar $O(N)$ model in the dimension range $4<d<6$ and study the effects caused by its metastability. As shown in previous work, this model formally possesses a fixed point where, perturbatively in the $1/N$ expansion, the operator scaling dimensions are real and above the unitarity bound.  Here, we further show that these scaling dimensions do acquire small imaginary parts due to the instanton effects.  In $d$ dimensions and for large $N$, we find that they are of order $e^{-N f(d)}$, where, remarkably, the function $f(d)$ equals the sphere free energy of a conformal scalar in $d-2$ dimensions.  The non-perturbatively small imaginary parts also appear in other observables, such as the sphere free energy and two and three-point function coefficients, and we present some of their calculations. Therefore, at sufficiently large $N$, the $O(N)$ models in $4<d<6$ may be thought of as complex CFTs.  When $N$ is large enough for the imaginary parts to be numerically negligible, the five-dimensional $O(N)$ models may be studied using the techniques of numerical bootstrap.  
}

\date{\centerline{\it Dedicated to the memory of Steve Gubser}}

\maketitle

\tableofcontents

\section{Introduction and summary}

One of the classic models of Quantum Field Theory (QFT) is the $O(N)$-symmetric theory of $N$ real scalar fields $\phi^i$ with interaction ${g\over 4} (\phi^i \phi^i)^2$. For small values of $N$ there are physical systems in three and two spacetime dimensions, 
whose critical behavior is described by this QFT\@.
While the infrared (IR) dynamics in these dimensions is strongly coupled, there are various methods for studying it, including dimensional continuation \cite{Wilson:1971dc,Wilson:1973jj} and the
$1/N$ expansion (for a review, see \cite{Moshe:2003xn}). Furthermore, the methods of conformal bootstrap \cite{Polyakov:1974gs,Ferrara:1973yt,Rattazzi:2008pe} 
(for reviews, see \cite{Rychkov:2016iqz,Simmons-Duffin:2016gjk,Poland:2018epd,Chester:2019wfx,Qualls:2015qjb}) 
have been fruitfully applied to the $O(N)$ model in various dimensions 
\cite{Kos:2013tga,Kos:2015mba,Chester:2014gqa,Li:2016wdp}. 

The $1/N$ expansion may be developed in continuous dimension $d$
using a generalized Hubbard-Stratonovich transformation with an auxiliary field $\sigma$ \cite{Vasiliev:1981yc,Vasiliev:1981dg,Vasiliev:1982dc,Lang:1990ni, Lang:1991kp, Lang:1992pp, Lang:1992zw, Petkou:1994ad,Petkou:1995vu}. For $N\geq 3$, the lower critical dimension is $2$, and the perturbative expansions in $d=2+\epsilon$ dimensions may be developed \cite{Brezin:1975sq} using the ultraviolet (UV) fixed point of
the $O(N)$ non-linear sigma model. (Recently, interesting non-local generalizations of the 2-d sigma model are being explored \cite{Gubser:2019uyf}.) 

In $d>4$, the quartic $O(N)$ model is non-renormalizable, but there is a UV fixed point in $d=4+\epsilon$ for $g<0$. Furthermore, the $1/N$ expansion may be formally continued to $d>4$. There is an interesting range, $4< d <6$, where the theory appears to be unitary order by order in the $1/N$ expansion \cite{Parisi:1975im,parisi1977non,Bekaert:2011cu,Bekaert:2012ux,Fei:2014yja}. 
Yet, the fate of the theory with large but finite $N$ is unclear in view of the expectation, supported by rigorous results \cite{Aizenman:1981du}, that the interacting $\phi^4$ theory cannot exhibit true critical behavior in $d>4$. 

Some light on this issue was shed by the series of papers starting with \cite{Fei:2014yja}, where a cubic $O(N)$-symmetric theory with the action for $N+1$ scalar fields given by
 \begin{equation}
S=\int d^d x \left(\frac{1}{2}(\partial \phi^i)^2+ \frac{1}{2}(\partial \sigma)^2+  \frac{g_1}{2}\sigma \phi^i \phi^i +\frac{g_2}{6}\sigma^3\right) \ ,
\label{Cubic6d}
\end{equation}
was introduced as a possible UV completion of the $O(N)$ model in $d<6$. Indeed, 
for $N> N_{\rm crit}$, where $N_{\rm crit}\approx 1038$, an IR stable fixed point of the theory (\ref{Cubic6d}) was found perturbatively in $\epsilon$
\cite{Fei:2014yja,Fei:2014xta,Gracey:2015tta}.
Furthermore, it was found that the $6-\epsilon$ expansions of various observables agree with the results obtained
from the formal $1/N$ expansions. For $N= N_{\rm crit}$ the IR fixed point merges with another fixed point, and for $N< N_{\rm crit}$ these fixed points become complex. This kind of merger of fixed points is a ubiquitous phenomenon in studies of the Renormalization Group \cite{Dymarsky:2005uh,Pomoni:2008de,Kaplan:2009kr,Giombi:2015haa,Grabner:2017pgm,Gorbenko:2018ncu,Gorbenko:2018dtm,Benini:2019dfy}, and theories at the complex fixed points for $N< N_{\rm crit}$ have been recently called ``complex CFTs" \cite{Gorbenko:2018ncu,Gorbenko:2018dtm}. 

\begin{figure}[!htb]
\begin{center}
  \includegraphics[width=10cm]{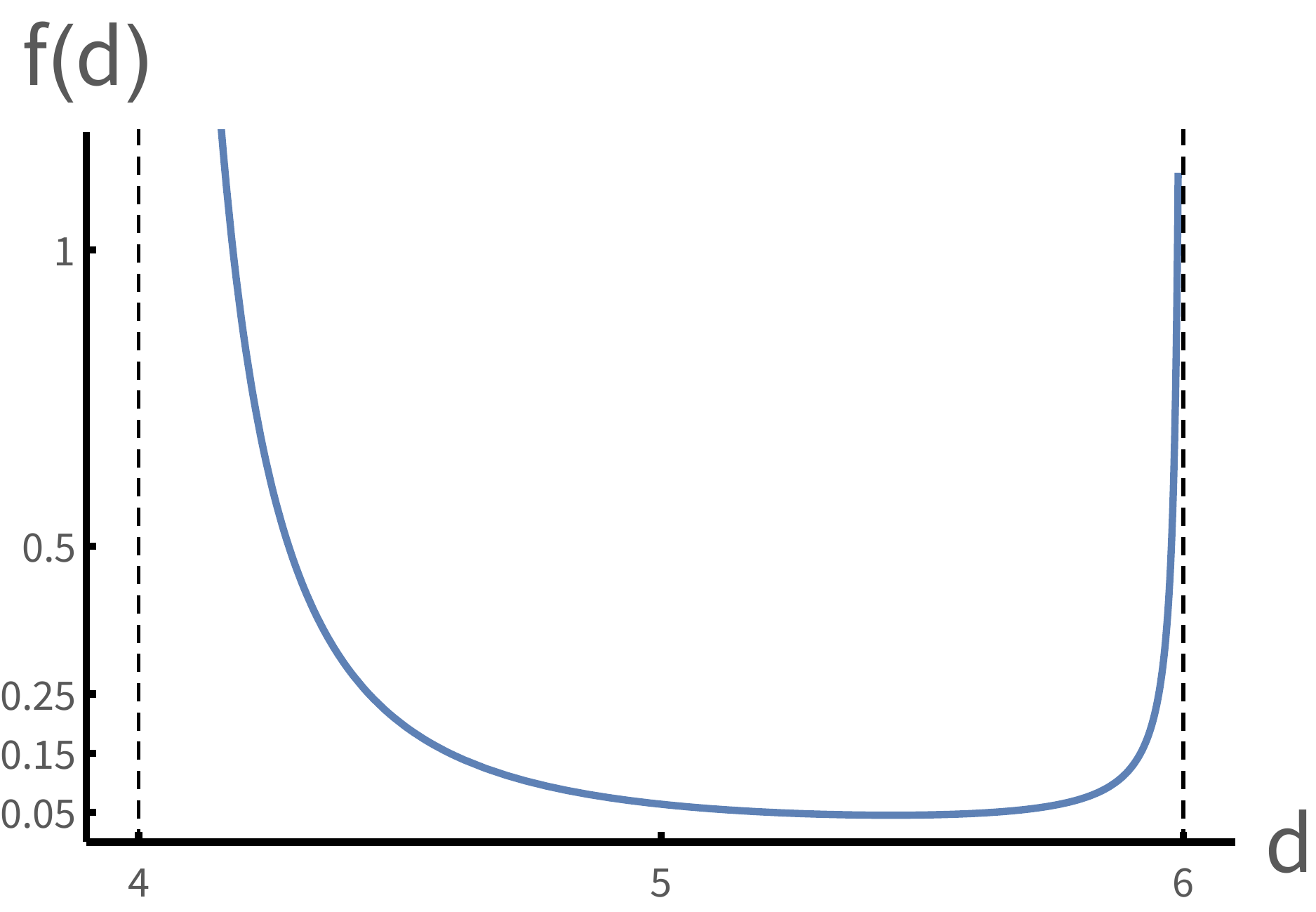} 
\caption{The function $f(d)$ controlling the non-perturbative imaginary parts $\sim \pm i e^{-N f(d)}$ in the large $N$ theory, plotted in the relevant range $4<d<6$, where it is positive. The function blows up near $d=4$ and $d=6$, corresponding to the fact that the $O(N)$ complex CFT becomes a free theory in those dimensions.}
\label{fd-plot}
\end{center}
\end{figure}

At the same time, one should expect that, even the model with $N> N_{\rm crit}$ cannot be perfectly stable because of tunneling from the perturbative vacuum at
 $\sigma=\phi^i=0$
to large negative values of $\sigma$, where the potential is unbounded from below. The instantons mediating this tunneling were found long ago in the 6-d cubic theory of a single scalar field 
\cite{Mckane:1978me}, as well as in the 4-d $O(N)$ model with negative coupling \cite{McKane:1978md,McKane:1984eq}. Via application of the instanton methods to the UV fixed point in 
$4+\epsilon$ dimensions, it was shown \cite{McKane:1984eq} that the critical exponents acquire imaginary parts of order $\exp \left (-\frac{N+8}{3\epsilon}\right )$.
 Extending the instanton calculations to the $O(N)$-symmetric model \eqref{Cubic6d} in $d = 6 - \epsilon$ dimensions, as well as to the large $N$ model in the range $4<d<6$, we find that various observables pick up imaginary parts that are exponentially suppressed as $e^{-N f(d)}$ at large $N$.\footnote{Similar large $N$ calculations have been performed on $\R^{d-1}\times S^1$ producing a thermal mass whose imaginary part 
in $d=5$ is not suppressed \cite{Petkou:2018ynm}. We reproduce these results using different methods in Appendix~\ref{thermal-App}.}   We calculate $f(d)$ and find that it is given by the free energy of a conformal scalar on $S^{d-2}$, which has the integral representation \cite{Giombi:2014xxa}: 
\begin{equation}
f(d)= \frac{1}{\sin(\pi d /2) \Gamma(d-1)}\int_0^1 dx x \sin(\pi x) \Gamma\left(\frac{d}{2}+x-1\right)\Gamma \left(\frac{d}{2}-x-1\right)\ .
\end{equation}
In particular, $f(5)=\frac {\log 2}{8} - \frac{3 \zeta(3)}{16 \pi^2} \approx 0.0638$ \cite{Klebanov:2011gs}.
The function $f(d)$ is plotted in Figure \ref{fd-plot}.
In agreement with \cite{McKane:1984eq}, 
$f(4+\epsilon)$ blows up as $1/(3\epsilon)$. Similarly, near six dimensions we find $f(6-\epsilon) \rightarrow 1/(90\eps)$; as we show below, this precisely agrees with the contribution of the instanton in the cubic theory (\ref{Cubic6d}). The relation between the non-perturbative imaginary parts in $d$ dimensions and the sphere free energy of a conformal scalar in $d-2$ dimensions is quite striking, and it would be nice to understand its physical origin.  
  
A simple argument for the exponential suppression of the imaginary parts is that
on the sphere $S^d$, or cylinder $S^{d-1}\times \R$, the conformal coupling to curvature adds a positive quadratic term to the scalar potential, making the perturbative vacuum metastable.
We demonstrate the smallness of imaginary parts explicitly by computing, at large $N$, the sphere free energy and the scaling dimensions of the operator $\phi^i$, which transforms as a vector of $O(N)$, and of the singlet $\sigma \sim \phi^i\phi^i$. We also show that other CFT data, such as three-point function coefficients and the normalization $C_J$ of the two-point function of the conserved $O(N)$ current, acquire the non-perturbatively small imaginary parts.\footnote{Technically similar calculations of instanton corrections to CFT correlation functions and operator scaling dimensions in the ${\cal N}=4$ SYM theory have been performed in \cite{Bianchi:1998nk, Dorey:1998xe, Dorey:1999pd, Green:2002vf, Kovacs:2003rt}. In that case there is no instability, and the instanton does not have a negative mode responsible for the imaginary parts.} 

The fact that the imaginary parts are very small for large $N$, makes the $O(N)$ models in $4<d<6$ similar to the robust examples \cite{Dymarsky:2005uh,Pomoni:2008de,Kaplan:2009kr,Giombi:2015haa,Grabner:2017pgm,Gorbenko:2018ncu,Gorbenko:2018dtm,Benini:2019dfy} of complex CFTs corresponding to the walking RG flows and weakly first-order phase transitions.\footnote{We should note, however, that the generation of small imaginary parts via the non-perturbative instanton effects is a different phenomenon than the merger and annihilation of perturbative fixed points. It would be interesting to look for other examples of complex CFTs where the small imaginary parts are generated non-perturbatively.} In $d=5$,  for large enough $N$ the imaginary parts of scaling dimensions can be made so small that the numerical bootstrap studies cannot distinguish such complex CFTs from the regular CFTs. This is probably the reason for the appearance in the $d=5$ conformal bootstrap of the islands in parameter space surrounding the values of operator dimensions which are in good agreement with the large $N$ expansions \cite{Li:2016wdp}. 

Our results may also be used to make predictions about the behavior of the $O(N)$ magnets on a $5$-dimensional lattice. While such spin systems are not expected to have non-gaussian second-order phase transitions, they may exhibit effects similar to near-criticality in a long-lived metastable state.  
Since the interacting $O(N)$ models in $4<d<6$ have {\it two} $O(N)$ invariant relevant operators \cite{Fei:2014yja,Fei:2014xta}, one may be able to find the approximately critical behavior by tuning both the nearest-neighbor and next-to-nearest-neighbor couplings on a lattice.\footnote{We are grateful to Slava Rychkov for this suggestion.} 

Our results may have interesting implications for the AdS/CFT correspondence \cite{Maldacena:1997re,Gubser:1998bc,Witten:1998qj}, 
in particular its higher spin version (for reviews see \cite{Giombi:2012ms,Giombi:2016ejx}). 
The type A
Vasiliev higher spin theories \cite{Vasiliev:1990en,Vasiliev:1992av, Vasiliev:2003ev} have the minimal field content, which consists of massless higher spin fields of even spin and a massive scalar. 
It has been conjectured \cite{Klebanov:2002ja} that such a theory in dimension $d+1$ is dual to the singlet sector of the $d$-dimensional $O(N)$ model, which is either free or interacting  depending on the choice of boundary conditions on the bulk scalar field. 
In $d=3$ both choices of boundary conditions produce a stable theory, since both 
free and interacting $O(N)$ models are conventional CFTs. In $d=5$, however, only the choice of boundary conditions corresponding to the free $O(N)$ model should be stable.
In view of our findings in this paper, the other choice is expected to be metastable, with the decay amplitude $\sim e^{-{\rm const}/G}$, where $G\sim 1/N$ is the bulk coupling constant. 
From the 
bulk point of view, the instability may again be due to instantons, whose existence should depend on the choice of boundary conditions.\footnote{Similar instanton solutions in $AdS_{d+1}$ were discussed in \cite{deHaro:2006ymc,Papadimitriou:2007sj}.}
It would be interesting to search for such instanton solutions of the higher-spin theory in $AdS_6$. 

The rest of this paper is organized as follows.  We begin in Section~\ref{CLASSICAL} with a description of the instanton solutions in the $d = 6 - \epsilon$ and $d = 4 + \epsilon$ expansions, where such solutions can be found by solving the classical equations of motion of the corresponding theories.  In Section~\ref{largeN-inst} we then describe the instantons at large $N$, where these instantons extremize the effective action for the Hubbard-Stratonovich field $\sigma$ that is obtained after integrating out the $O(N)$ vector fields $\phi^i$. These solutions on $S^d$ have constant $\sigma=-k(k+1)$ (for a special choice of the instanton moduli), where $k$ is a positive integer. In $4<d<6$ the dominant non-perturbative effects come from the $k=1$ saddle point. We continue with calculations of the instanton contribution to the round sphere free free energy in Section~\ref{F-Sd}, to the operator scaling dimensions in Section~\ref{SCALING}, and to $C_J$ and an example of a three-point function coefficient in Section~\ref{COEFFICIENTS}.  In Section~\ref{OTHERSADDLES} we
exhibit the classical solutions on $S^4$ and $S^6$
 where the fields $\phi^i$ are not constant, but are rather proportional to spherical harmonics; these solutions correspond to the large $N$ saddle points with $k>1$.  
Several technical details and related calculations are relegated to the Appendices.
 
\section{Classical instantons in the epsilon expansion}
\label{CLASSICAL}

\subsection{The instanton near six dimensions}
\label{Inst-6d}
The equations of motion of the $O(N)$ invariant cubic scalar theory with action (\ref{Cubic6d}) 
are given by
\begin{equation}
\begin{aligned}
\nabla^2 \phi^i &= g_1 \sigma \phi^i \,, \\
\nabla^2 \sigma &= \frac{g_1}{2} \phi^i\phi^i + \frac{g_2}{2} \sigma^2\,.
\label{Class-EOM}
\end{aligned}
\end{equation}
In $d=6$, in addition to the trivial solution $\sigma = \phi^i = 0$, these equations admit the $O(N)$-invariant instanton solution
\begin{align}
\phi^i &=0 \,, \\
\sigma &= -\frac{12}{g_2} \frac{4\lambda^2}{(1+\lambda^2 (\vec{x}-\vec{a})^2)^2}\,. \label{sigmac}
\end{align} 
Here $\vec{a}$ and $\lambda$ are the moduli corresponding to the position and size of the instanton. Since this solution has $\phi^i=0$, it is a simple generalization of the instanton solution in the single scalar cubic theory studied in \cite{Mckane:1978me}.\footnote{In the theory with classical action \eqref{Cubic6d} there are other instanton-like solutions, where both $\sigma$ and $\phi^i$ are non-zero.  We will discuss these classical solutions in Section~\ref{OTHERSADDLES}. }  Plugging this solution into \eqref{Cubic6d}, one obtains the finite instanton action 
\begin{equation}
S_{\rm inst} = \frac{768\pi^3}{5 g_2^2}\,.
\label{instAct-6d}
\end{equation}

The instanton solution (\ref{sigmac}) is responsible for tunneling from the metastable ground state at $\sigma=\phi^i=0$. 
As we show explicitly in Appendix~\ref{det-6d}, computing the spectrum of quantum fluctuations around this solution, one finds a single negative mode, so that the instanton yields an imaginary contribution $\sim \pm i e^{-S_{\rm inst}}$ to the free energy and to other observables. To leading order in the $d=6-\epsilon$ expansion, for $N>N_{\rm crit}\approx 1048$ one finds a fixed point at \cite{Fei:2014yja}
 \es{gstar}{
  g_1^* &= \sqrt{\frac{6\epsilon(4\pi)^3}{N}}\left(1+\frac{22}{N}+\ldots\right)\,, \\
  g_2^* &= 6\sqrt{\frac{6\epsilon(4\pi)^3}{N}}\left(1+\frac{162}{N}+\ldots\right) \,, 
 }
so that at large $N$ and small $\epsilon$ the instanton action is
\begin{equation}
S_{\rm inst}^{6-\epsilon} = \frac{N}{90\epsilon}\left(1+O(1/N)\right)\,.
\label{Sinst-eps}
\end{equation}
Hence, the imaginary contribution to free energy and other observables comes with the exponentially suppressed factor $\sim \pm i e^{-\frac{N}{90\epsilon}}$. Let us point out that the $\beta$-function of the theory is also expected to receive such imaginary contributions from the instanton (see \cite{McKane:1978md, Mckane:1978me, McKane:1984eq}), so that the value of the fixed point couplings (\ref{gstar}) gets small imaginary parts as well. This would give non-perturbative corrections to the action (\ref{Sinst-eps}), which we neglect in this section as they are further suppressed. 

Let us now consider a conformal mapping of the theory (\ref{Cubic6d}) from flat space to the unit-radius round sphere $S^d$ parameterized in stereographic coordinates $\vec{x}$, with the metric
\begin{equation} 
ds_{S^d}^2 = \frac{4d\vec{x}^2}{(1 + \vec{x}^2)^2} \,.
\label{Sd-metric}
\end{equation}
In $d=6$, the action (\ref{Cubic6d}) and the equations of motion (\ref{Class-EOM}) are conformally invariant, so the instanton solution on $S^6$ can be simply obtained by a Weyl rescaling, and it is given by
\begin{equation}
\sigma = -\frac{12}{g_2} \frac{\lambda^2(1 + \vec{x}^2)^2}{(1+\lambda^2 (\vec{x}-\vec{a})^2)^2}\,,
\label{inst-S6}
\end{equation}
where we have used that $\sigma$ has scaling dimension 2. 
It is easy to check that the classical action on $S^6$
 \es{S6-action}{
S = \int d^6x\sqrt{g}\left(\frac{1}{2}\partial^{\mu}\phi^i\partial_{\mu}\phi^i 
+\frac{1}{2}\partial^{\mu}\sigma\partial_{\mu}\sigma 
+3\sigma^2+3 \phi^i\phi^i 
+ \frac{g_1}{2}\sigma\phi^i \phi^i+\frac{g_2}{6}\sigma^3 \right)
 }
evaluated on the solution (\ref{inst-S6}) has the same value as (\ref{instAct-6d}), as expected from conformal invariance. The quadratic terms in the action above come from the conformal coupling to the sphere curvature.\footnote{In general $d$, the contribution of the conformal coupling to the Lagrangian for a scalar field $\phi$ is given by $\frac{d-2}{8(d-1)}{\cal R}\phi^2$, where ${\cal R}$ is the Ricci scalar. On $S^d$, we have ${\cal R}=d(d-1)/R^2$, and we have set $R=1$ in (\ref{S6-action}).}
\begin{figure}[!htb]
\begin{center}
  \includegraphics[width=8cm]{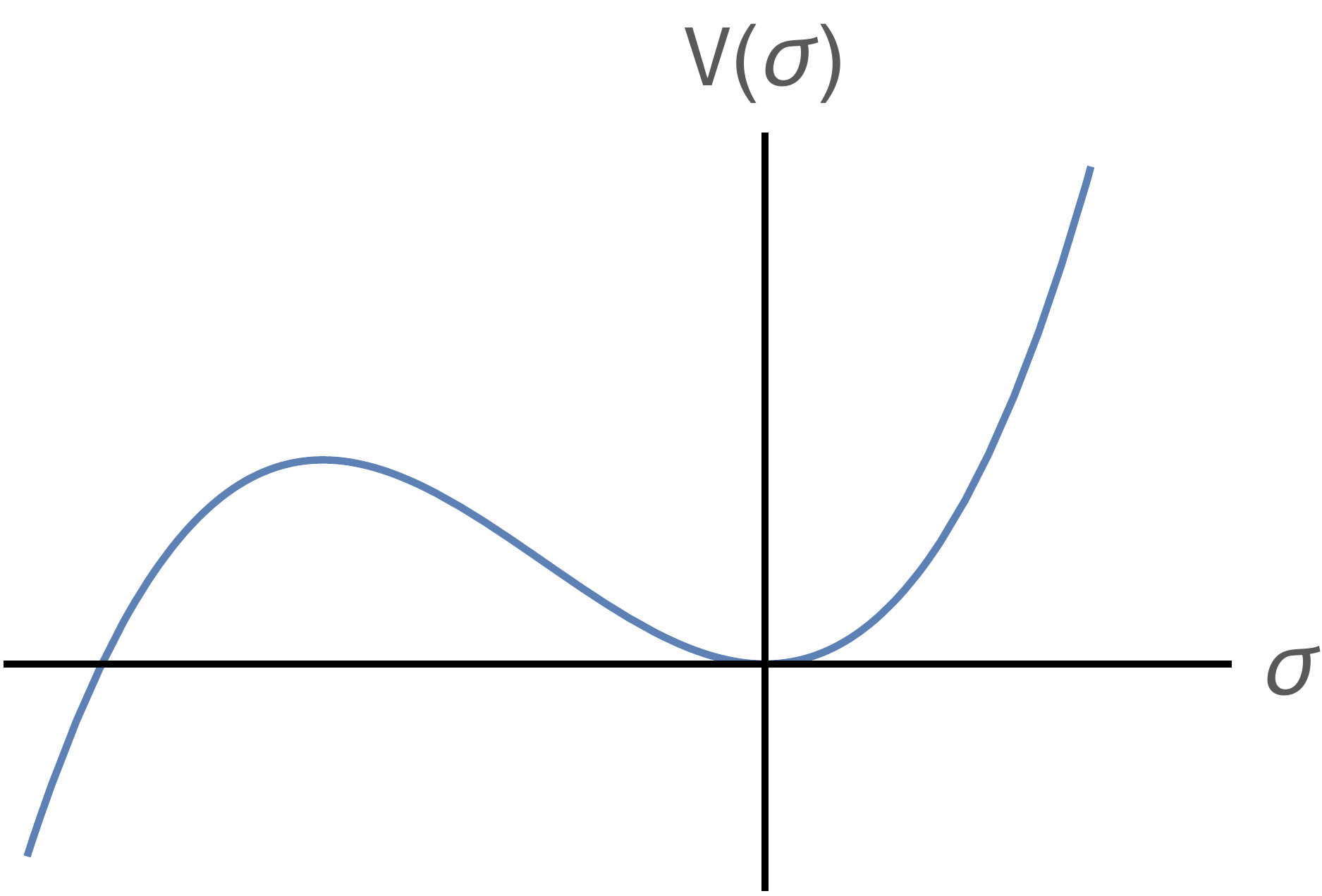} 
\caption{Potential for constant $\sigma$ for the cubic $O(N)$ theory on $S^6$. The instanton solution (\ref{sigma-S6}) corresponds to the local maximum of the potential.}
\label{fig:pot6d}
\end{center}
\end{figure}

We now observe that for a special choice of the moduli, $\lambda=1$\footnote{Or $\lambda=1/R^2$ if we reinstate the radius of the sphere.} and $\vec{a}=0$, the instanton solution on the sphere is just a constant\footnote{This was also noticed and used in related models.  For example, see \cite{McKane:1978md, Drummond:1978pf, Mckane:1978me}.}
\begin{equation}
\sigma = -\frac{12}{g_2}\,.
\label{sigma-S6}
\end{equation}
This has a simple interpretation: it is the critical point corresponding to the local maximum of the potential for configurations with constant $\sigma$ on $S^6$:
\begin{equation}
V(\sigma) = 3\sigma^2+\frac{g_2}{6}\sigma^3\,.
\label{V6d}
\end{equation}
A plot of this potential is given in Figure \ref{fig:pot6d}. 

Note that one may also conformally map the theory to the cylinder $\mathbb{R}_t \times S^{5}$. The constant solution (\ref{sigma-S6}) is then mapped to the time-dependent configuration
\begin{equation}
\sigma = -\frac{12}{g_2 \cosh^2t}\,.
\label{R-S-sol}
\end{equation}
This solves the equation 
\begin{equation}
\partial_t^2\sigma = 4\sigma+\frac{g_2}{2}\sigma^2\,,
\end{equation}
where the right-hand side comes from the potential on $\mathbb{R}_t\times S^{5}$, $V_{\R_t \times S^{5}} = 2\sigma^2+g_2\sigma^3/6$. The solution (\ref{R-S-sol}) has just the same form as the instanton in quantum mechanics which is responsible for tunneling in a cubic potential. See, for instance, \cite{ZinnJustin:2002ru, Marino:2015yie} for reviews. In the rest of the paper we will focus mainly on the $S^d$ description of the instanton.

In section \ref{largeN-inst}, we will describe how an analogous instanton solution arises in the large $N$ treatment of the $O(N)$ model in $4<d<6$ (the integer dimension $d=5$ is, of course, the most interesting). Essentially, the role of the ``fundamental" field $\sigma$ above will be played by the Hubbard-Stratonovich field, which is used to develop the large $N$ expansion of the model in general $d$. 

\subsection{The instanton near four dimensions}
\label{Inst-4d}
For completeness, let us also discuss how the instanton solution looks like near the lower end of the range $4<d<6$, where we can formally use the $d=4+\epsilon$ expansion in the $O(N)$ invariant quartic scalar theory 
\begin{equation}
S=\int d^d x \left(\frac{1}{2}(\partial \phi^i)^2+\frac{g}{4}(\phi^i \phi^i)^2\right)\,.
\label{ON4d}
\end{equation}
In $d=4+\epsilon$, the one-loop beta function is given by
\begin{equation}
\beta_g = \epsilon g + \frac{N+8}{8\pi^2}g^2 \,,
\end{equation}
and thus we see that there is a formal UV fixed point at negative coupling
 \es{gstar4}{
g_* = -\frac{8\pi^2\epsilon}{N+8}+O(\epsilon^2)\,.
 }
At the level of perturbation theory in $\epsilon$, the corresponding fixed point appears to be unitary (all scaling dimensions are real and above the unitarity bound) to all orders in $\epsilon$. However, due to the wrong sign quartic potential, we expect the model to be non-perturbatively unstable. Indeed, it is well known that for negative coupling the theory (\ref{ON4d}) in $d=4$ has a real instanton solution \cite{Brezin:1976wa, McKane:1978md}
\begin{equation}
\phi^i = \sqrt{-\frac{8}{g}} \frac{\lambda}{1+\lambda^2(\vec{x}-\vec{a})^2} \hat{u}^i 
\end{equation}
where $\hat{u}^i$ is a constant unit $N$-component vector ($\hat{u}^i\hat{u}^i=1$), and $\lambda, \vec{a}$ are the size and position moduli as in the previous section. Note that $\hat{u}^i$ are also exact moduli parameterizing an $S^{N-1}$. Integration over these moduli restores $O(N)$ invariance of correlation functions (see for instance \cite{ZinnJustin:2002ru} for a review). 

\begin{figure}[!htb]
\begin{center}
\includegraphics[width=8cm]{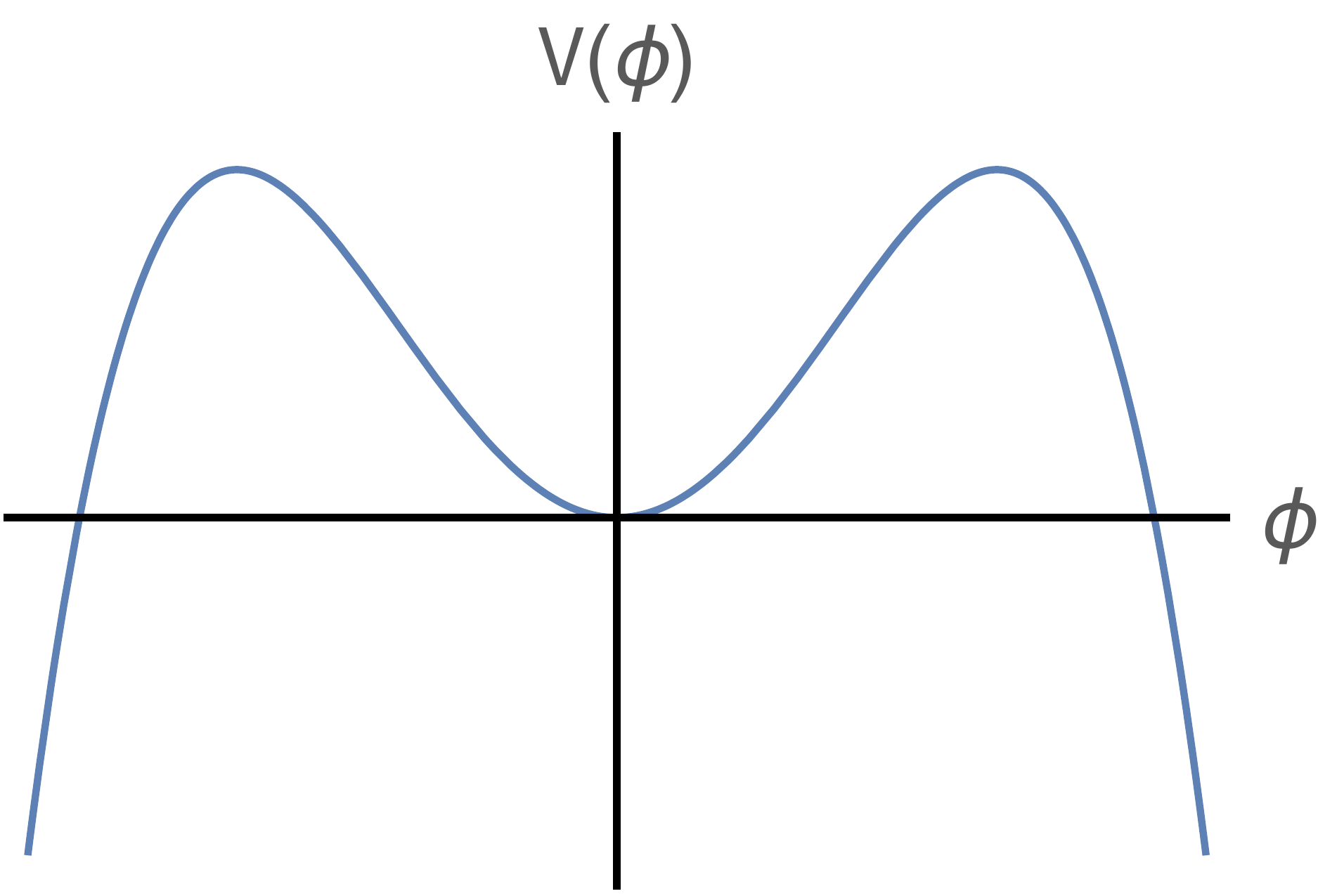} 
\caption{Potential for $\phi$ in the quartic theory on $S^4$ for negative coupling, depicted here in the case of $N=1$. The instanton solution (\ref{inst-4d-const}) corresponds to the local maxima of the potential (for $N=1$, one has $\hat{u}=\pm 1$).}
\label{fig:pot4d}
\end{center}
\end{figure}

After a conformal mapping to $S^4$, the instanton solution takes the form
\begin{equation}
\phi^i = \sqrt{-\frac{2}{g}} \frac{\lambda(1+\vec{x}^2)}{1+\lambda^2(\vec{x}-\vec{a})^2} \hat{u}^i 
\end{equation}
which solves the equations of motion of the $S^4$ theory with the action
\begin{equation}
S = \int d^4x\sqrt{g}\left(\frac{1}{2}\partial^{\mu}\phi^i\partial_{\mu}\phi^i 
+\phi^i\phi^i 
+ \frac{g}{4}(\phi^i \phi^i)^2 \right)\,.
\label{S4-action}
\end{equation}
Again, we observe that on the sphere there is a choice of moduli where the solution becomes a constant
\begin{equation}
\phi^i = \sqrt{-\frac{2}{g}} \hat{u}^i\,,
\label{inst-4d-const}
\end{equation}
which can be seen to correspond to the degenerate maxima of the potential
\begin{equation}
V(\phi)= \phi^i\phi^i + \frac{g}{4}(\phi^i \phi^i)^2\,.
\end{equation}
As discussed above, the conformal coupling to the sphere curvature makes the perturbative vacuum metastable for $g<0$, as shown in Figure \ref{fig:pot4d}. 
The value of the classical action on the instanton solution (computed either on $\mathbb{R}^4$ or $S^4$) is
\begin{equation}
S_{\rm inst}=-\frac{8\pi^2}{3g} \,.
\end{equation}
Plugging in the value of the coupling at the fixed point in $d=4+\epsilon$, this yields
\begin{equation}
S_{\rm inst}^{4+\epsilon}=\frac{N+8}{3\epsilon}\, ,
\label{Sinst-4d}
\end{equation}
in agreement with \cite{McKane:1984eq}.
We will see that this matches with the large $N$ method we introduce in the next section. In Section~\ref{OTHERSADDLES} we show that there are additional classical solutions where
$\phi^i$ is generalized from a constant to any spherical harmonic on $S^d$.

\section{``Hubbard-Stratonovich instanton" at large $N$}
\label{largeN-inst}

Applying the familiar Hubbard-Stratonovich transformation to the $(\phi^i\phi^i)^2$ interaction in the $O(N)$ model, one arrives at the following action describing the critical $O(N)$ model:
\begin{equation}
S_{\rm crit}=\int d^d x \left(\frac{1}{2}(\partial \phi^i)^2+\frac{1}{2}\sigma \phi^i \phi^i \right) \,, \label{LargeNaction}
\end{equation}
where $\sigma$ is the Hubbard-Stratonovich field, and we have dropped the term $\sim \sigma^2/\lambda$ which becomes irrelevant in the critical limit.\footnote{We will use the same symbol $\sigma$ as in the previous section to denote this field, although of course the two quantities are not identical, as they have different normalizations.} 
This action can be used to systematically develop the $1/N$ expansion of the theory. The field $\sigma$, which acquires induced dynamics due to $\phi$ loops, becomes at large $N$ a conformal scalar operator with scaling dimension $\Delta=2+O(1/N)$. 

After a conformal transformation to the sphere metric (\ref{Sd-metric}), one obtains the action
\begin{equation}
S_{\rm crit}=\int d^d x \sqrt{g}\left(\frac{1}{2}\partial^{\mu}\phi^i\partial_{\mu}\phi^i +\frac{d(d-2)}{8}\phi^i \phi^i+\frac{1}{2}\sigma \phi^i \phi^i \right)\,.\label{LargeNaction-Sd}
\end{equation}
Since this action is quadratic in $\phi^i$, we can integrate out these fields exactly, and hence obtain a path integral over $\sigma$ with action
\begin{equation}
S_{\sigma} = \frac{N}{2}\log \det \left(-\nabla^2 + \frac{d(d-2)}{4}+\sigma \right)\,.
\end{equation}
Given the intuition from the classical solution in $d=6$ described in the previous section, it is natural to look for configurations of constant $\sigma$ that solve the equation of motion. Recall that the eigenvalues $\lambda_n$ and degeneracies $D_n$ of the scalar laplacian on $S^d$ are:
 \es{lamDLap}{
\lambda_n &= n(n+d-1) \,, \\
 D_n &= \frac{(2n+d-1)\Gamma(n+d-1)}{n!\Gamma(d)}\,.
 }
So the action for constant $\sigma$ is
\begin{equation}
S_{\sigma}= \frac{N}{2}\sum_{n=0}^{\infty} D_n \log \left(n(n+d-1)+\frac{1}{4}d(d-2)+\sigma \right) \,.
\label{Ssig-class}
\end{equation}
Following \cite{Giombi:2014xxa}, we find
\begin{equation}
\frac{\partial S_{\sigma}}{\partial \sigma} = \frac{N}{2\sin(\frac{\pi d}{2})\Gamma(d)}\Gamma\left(\frac{d-1}{2}+i\sqrt{\sigma-\frac{1}{4}} \right)\Gamma\left(\frac{d-1}{2}-i\sqrt{\sigma-\frac{1}{4}} \right)\cosh \left(\pi \sqrt{\sigma-\frac{1}{4}} \right) \,.
\end{equation}
Consequently, the action \eqref{Ssig-class} is extremized when
\begin{equation}
\sqrt{\sigma-\frac{1}{4}}=i\frac{2k+1}{2}
\end{equation}
for some integer $k$. This gives
\begin{equation}
\sigma = -k(k+1)\,.
\end{equation}
The $k=0$ solution is the perturbative vacuum where the $N$ fields $\phi^i$ are conformally coupled scalars. From (\ref{Ssig-class}) we see that $S_{\sigma}$ diverges logarithmically when $\sigma$ approaches the value $\sigma=-d(d-2)/4$. This is just the point where the effective mass of $\phi^i$, i.e. $m^2=d(d-2)/4+\sigma$, goes to zero. 
For $d<4$, one can see that only the perturbative saddle $k=0$ lies to the right of this divergence. On the other hand, for $4<d<6$, both the $k=0$ and $k=1$ saddles lie in the region $\sigma>-d(d-2)/4$.\footnote{For $d>6$, additional solutions lie
in the region $\sigma>-d(d-2)/4$ (e.g. for $6<d<8$, we can have $k=0,1,2$), and it would be interesting to understand their physical interpretation; we leave this to future work.} It seems natural to assume that an appropriate integration contour in the $\sigma$ complex plane can be chosen so that the leading contributions at large $N$ come from the saddle points with $\sigma>-d(d-2)/4$ ($k=0,1$ for $4<d<6$), as the physical interpretation of the additional saddles seems unclear (in particular, as will be shown in Section \ref{OTHERSADDLES} below, the solutions with $k>1$ have several negative modes). For instance, for $d<4$ the appropriate contour can be chosen to run along the imaginary $\sigma$ axis passing only through the $k=0$ saddle.  

\begin{figure}[htbp]
\begin{center}
  \includegraphics[width=9cm]{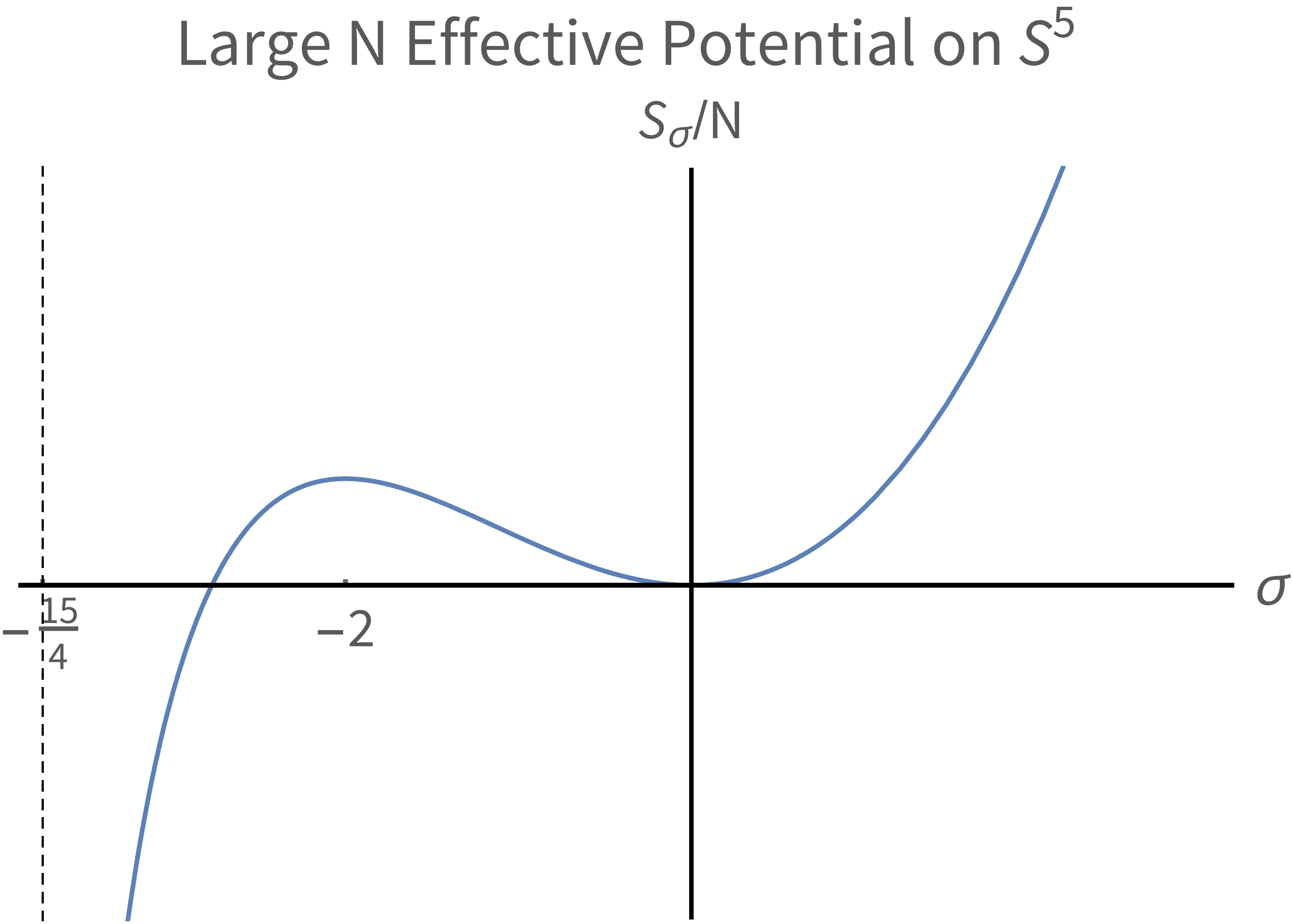} 
\caption{Effective potential for constant $\sigma$ in the large $N$ theory on $S^5$. The instanton solution $\sigma=-2$ corresponds to the local maximum of the potential.}
\label{fig:potS5}
\end{center}
\end{figure}

Let us now focus on the $k=1$ solution in $4<d<6$, i.e.~that with $\sigma=-2$. As in the 
cubic theory in $d=6$, this solution can be seen to correspond to the local maximum of the effective potential for constant $\sigma$ in Eq.~(\ref{Ssig-class}). A plot of this potential in $d=5$ is given in Figure \ref{fig:potS5} (for general $4<d<6$, the behavior is analogous); it is qualitatively similar to the classical potential for the cubic theory on $S^6$ (except that it becomes unbounded from below as $\sigma$ approaches $-15/4$ as explained above, while in $d=6-\epsilon$ this happens only asymptotically as $\sigma \rightarrow -\infty$). It is then natural to view the $\sigma=-2$ solution as the instanton configuration that is the large $N$ counterpart of the classical solution described in the previous section. Indeed, as we will show below, studying the spectrum of fluctuations around then $\sigma=-2$ saddle point, one finds a single negative mode and $d+1$ zero modes, which we interpret as the size and position moduli of the instanton. Hence, even though we have found the solution specializing to constant configurations of $\sigma$, we expect that it belongs to a family of instanton solutions
\begin{equation}
\sigma = -2 \frac{\lambda^2(1 + \vec{x}^2)^2}{(1+\lambda^2 (\vec{x}-\vec{a})^2)^2}\,,
\label{inst-Sd}
\end{equation}
where $\lambda$ and $\vec{a}$ are the moduli, as seen explicitly in the $d=6$ analysis. We will compute the value of the action (\ref{Ssig-class}) on the instanton solution, as well as the determinant for quantum fluctuations, in Section \ref{F-Sd} below. Before doing that, in the next section we give a useful description of the instanton profile using embedding coordinates in the instanton moduli space.  

\subsection{Instanton profile in embedding coordinates}
\label{Inst-Embed}

On the round $S^d$ with metric (\ref{Sd-metric}), the instanton profile in the large $N$ theory is given by (\ref{inst-Sd}). If we make a conformal transformation to flat $\R^d$ (parameterized by $\vec{x}$ and with line element $d\vec{x}^2$), the profile for $\sigma$ will then be
 \es{sigmaFlat}{
  \R^d: \qquad \sigma = \frac{-8 \lambda^2}{\left(1 + \lambda^2 (\vec{x} - \vec{a})^2 \right)^2}  \,.
 }

As familiar in instanton calculus, the moduli space is expected to be given by 
the quotient $SO(d+1,1)/SO(d+1)$, which is the $(d+1)$-dimensional hyperbolic space $\HH^{d+1}$.\footnote{This statement is clear when we study the theory on $S^d$, where the instanton solution preserves the $SO(d+1)$ rotational symmetry of $S^d$.  Since the effective action for $\sigma$ is classically conformally invariant, it is invariant under $SO(d+1, 1)$.  Thus, we expect the constant $\sigma$ solution to correspond to a point on the manifold of solutions $SO(d+1,1)/SO(d+1)$.}  We can then simplify the formulas by going to a $(d + 2)$-dimensional embedding space with signature $(+, +, \cdots, +, -)$. In this space, $\HH^{d+1}$ is given by the hyperboloid
 \es{Hyperboloid}{
  X \cdot X = \sum_{i=1}^{d+1} (X_i)^2 - X_{d+2}^2 = -1 \,.
 }
The relation between these coordinates and $\vec{a}$ and $\lambda$ is
 \es{XFromalam}{
  X = \begin{pmatrix} 
    \lambda \vec{a} \ ,\ - \frac{1}{2 \lambda} + \lambda \frac{1 - \vec{a}^2}{2}\ , \ \frac{1}{2 \lambda} + \lambda \frac{1 + \vec{a}^2}{2} 
    \end{pmatrix} \,.
 }
For the CFT coordinates, we use null vectors $P$ in the same $(d+2)$-dimensional space:
 \es{Null}{
   P \cdot P = \sum_{i=1}^{d+1} (P_i)^2 - P_{d+2}^2 = 0 \,.
 }
For flat space, we have
 \es{FlatP}{
   \R^d: \qquad P = \begin{pmatrix} 
     \vec{x} \ ,\  \frac{1 - \vec{x}^2}{2}\ , \  \frac{1 + \vec{x}^2}{2} 
     \end{pmatrix} \,, 
 }
while for the sphere we have
 \es{SphereP}{
   S^d: \qquad P = \begin{pmatrix} 
     \frac{2\vec{x}}{1 + \vec{x}^2}  \ ,\   \frac{1 - \vec{x}^2}{1 + \vec{x}^2}\ , \  1 
    \end{pmatrix} \,.
 }
These null vectors have the property that the induced line element $dP \cdot dP = \sum_{i=1}^{d+1} (dP_i)^2 - dP_{d+2}^2 $ is precisely equal to the metric on flat space and on the unit sphere, respectively.

In terms of these coordinates, the instanton profile (either on flat space or on the sphere) can be written as
 \es{InstProfile}{
  \sigma = - \frac{8}{(-2 X \cdot P )^2} \,.
 }

A different coordinate system that makes the symmetries of $S^d$ manifest can be found as follows.  We parameterize $S^d$ by a unit vector $\hat p$ embedded in $\R^{d+1}$, and we parameterize the instanton moduli space by a radial coordinate $\rho$ and a unit vector $\hat n$ in $\R^{d+1}$.  We take
 \es{PXParam}{
  P &= \begin{pmatrix} \hat p \ , \ 1 \end{pmatrix} \,, \\
  X &= \begin{pmatrix} \hat n \sinh \rho \ , \ \cosh \rho \end{pmatrix}  \,.
 }
To find the relation between $\hat p$ and $\vec{x}$ we equate the first equation in \eqref{PXParam} to \eqref{SphereP}, and to find the relation between $(\rho, \hat n)$ and $(\lambda, \vec{a})$ we should equate the second equation in \eqref{PXParam} to \eqref{XFromalam}.  In these new coordinates, the line elements on $S^d$ and on the instanton moduli space are 
 \es{Metrics}{
  dP \cdot dP = d\hat p^2 \,, \qquad
   dX \cdot dX = d \rho^2 + \sinh^2 \rho\,  d \hat n^2 \,.
 }
The instanton profile is
 \es{InstProfileAgain}{
  \sigma = - \frac{2}{(\cosh \rho - \sinh \rho\, \hat p \cdot \hat n )^2} \,.
 } 
Quite nicely, for $\rho \neq 0$, an $SO(d+1)$ rotation of the point $\hat p$ on $S^d$ is equivalent to a rotation of the unit vector $\hat n$.  If we want to preserve rotational symmetry on $S^d$ when integrating over the moduli space of instantons, we should thus impose a cutoff at a fixed value of $\rho$.\footnote{Note that at large $\rho$, the instanton profile \eqref{InstProfileAgain} becomes highly peaked around $\hat p = \hat n$.  If $\theta$ is the angle between $\hat p$ and $\hat n$ defined by $\cos \theta = \hat p \cdot \hat n$, then from \eqref{InstProfileAgain} it can be shown that the angular width $\Delta \theta$ of this peak scales as $\Delta \theta \sim e^{-\rho}$ at large $\rho$.   A cutoff $\rho_m$ at large $\rho$ is then equivalent to imposing a small angle cutoff equal to $e^{-\rho_m}$.  If $\epsilon$ is the short distance cutoff and $R$ is the radius of the sphere, we can thus identify $e^{\rho_m} = R/\epsilon$.}  We will make such a choice shortly when computing the instanton contribution to the sphere partition function and to various correlation functions.

\section{The $S^d$ partition function}
\label{F-Sd}

In the large $N$ expansion, we have that the $S^d$ partition function is (keeping up to order $N^0$ terms for each saddle):
 \es{ZSd}{
  Z_{S^d} &= A_0(d) e^{- N f_0(d)} +  \Vol(\HH^{d+1}) A_1(d) e^{-N f_1(d)} + \cdots \\
   &= A_0(d) e^{- N f_0(d)} \left[ 1 +\Vol(\HH^{d+1}) \frac{A_1(d)}{A_0(d)} e^{- N (f_1(d) - f_0(d))} + \cdots \right] \,.
 }
The quantities $f_0$ and $f_1$ come from the $\phi$ determinant, and may be viewed as the ``classical action" on the $\sigma$ saddle point, while $A_0$ and $A_1$ come from the $\sigma$ determinants for quadratic fluctuations around the saddle point.   In the contribution of the $\sigma$ determinant around the instanton saddle, we included an explicit factor of the (divergent) volume $\Vol(\HH^{d+1})$ of the unit curvature radius hyperbolic space, in anticipation of the fact that the integration over the instanton moduli space yields such a factor.

More explicitly, the partition function around a saddle point $\sigma_c$ to quadratic order in fluctuations is given by
 \es{ZExp}{
  &Z_{S^d}|_{\sigma_c} =\int D\sigma e^{-\frac{N}{2} \tr \log (-\nabla^{2}+ \frac{d(d-2)}{4}+\sigma)}|_{\sigma=\sigma_c+\delta \sigma} \\
    &\approx e^{-\frac{N}{2} \tr \log (-\nabla^{2}+ \frac{d(d-2)}{4}+\sigma_{c})} \int [D\delta \sigma] 
         e^{-\frac{N}{2}  \int d^{d}\vec{x}\,  d^{d}\vec{y}\,  \delta\sigma(\vec{x}) (-G^{2}(\vec{x},\vec{y})) \delta\sigma(\vec{y})}\\
    &= (\det (-NG^{2}))^{-1/2}e^{-\frac{N}{2} \tr \log (-\nabla^{2}+\frac{d(d-2)}{4}+ \sigma_{c})}, 
}
where $G(\vec{x},\vec{y})$ denotes the Green's function for $\phi$ in the background $\sigma_c$, which will be computed below. 

\subsection{``Classical action" from $\phi$ determinant}

Let us now calculate the quantities appearing in \eqref{ZSd}.  From (\ref{Ssig-class}), we find that for the  perturbative vacuum at  $\sigma=0$ we have
 \es{f0}{
  f_0(d) =  \frac 12 \sum_{n=0}^\infty D_n \log \left[ \left( n + \frac d2 \right) \left( n -1 + \frac{d}{2} \right)  \right] 
   = \frac 12 \sum_{n=0}^\infty (D_{n-1} + D_{n}) \log \left( n  - 1 + \frac d2 \right) \,.
 }
For the $\sigma = -2$ instanton solution, we instead have
 \es{f1}{
  f_1(d) = \frac 12 \sum_{n=0}^\infty D_n \log \left[ \left( n + 1 + \frac {d} 2 \right) \left( n - 2 + \frac{d}{2} \right)  \right] 
   = \frac 12 \sum_{n=0}^\infty  (D_{n-3} + D_{n})  \log \left( n - 2+ \frac d2 \right) \,.
 }
Subtracting, we obtain
 \es{f1mf0}{
  f_1(d) - f_0(d) = \frac 12 \sum_{n=0}^\infty
   \biggl[ D_{n-3} (d)+ D_{n} (d)- D_{n-2} (d)- D_{n-1}(d) \biggr] \log\left( n - 2+ \frac d2 \right) \,.
 }
It can be easily checked using a useful relation $D_{n}(d)=D_{n-1}(d)+D_{n}(d-1)$ that
 \es{f1mf0Again}{
  f_1(d) - f_0(d)  = \frac 12 \sum_{n=0}^\infty
   \biggl[ D_{n} (d-2)+ D_{n-1} (d-2) \biggr] \log\left( n - 1+ \frac {d-2}2 \right) \,.
 }
This expression is identical to \eqref{f0} with $d$ shifted down by $2$.  So we find
 \es{f1mf0Final}{
  f_1(d) - f_0(d) = f_0(d - 2)  = F_{S^{d-2}}^{\text{free scalar}} \,.
 } 
Since the derivation above involved some formal manipulations with divergent sums, in Appendix~\ref{oneloopdet} we present an alternative derivation of this result---see Eq.~(\ref{f-alpha}). 

In $d = 5$, we then have
\begin{equation}
f_1(5) - f_0(5) =F_{S^{3}}^{\text{free scalar}} =\frac {\log 2}{8} - \frac{3 \zeta(3)}{16 \pi^2} \approx 0.0638 \,,
\end{equation}
where we have used the value of $F_{S^{3}}^{\text{free scalar}}$ computed in \cite{Klebanov:2011gs}. 

In general $d$, the value of $F$ for a free conformal scalar has the integral representation \cite{Giombi:2014xxa}
\begin{equation}
\begin{aligned}
F_{S^{d}}^{\text{free scalar}} &= \frac{1}{2} \log \det \left(-\nabla^2 + \frac{d(d-2)}{4} \right) \\
&= -\frac{1}{\sin(\pi d /2) \Gamma(d+1)}\int_0^1 dx x \sin(\pi x) \Gamma(\frac{d}{2}+x)\Gamma(\frac{d}{2}-x) \,.
\label{Ffreesc}
\end{aligned}
\end{equation}
This expression can be easily expanded near even integer dimensions, where one finds poles related to the conformal $a$-anomaly coefficient. For the instanton in $d=6-\epsilon$ dimensions, we find from (\ref{f1mf0Final}) and using the above representation that
\begin{equation}
f_1(6-\epsilon) - f_0(6-\epsilon)= F_{S^{4-\epsilon}}^{\text{free scalar}}=\frac{1}{90\epsilon} \,,
\end{equation}
which is in precise agreement with the result (\ref{Sinst-eps}) obtained from the classical instanton in the 6-d cubic theory. In the case $d=4+\epsilon$, we find similarly
\begin{equation}
f_1(4+\epsilon) - f_0(4+\epsilon)= F_{S^{2+\epsilon}}^{\text{free scalar}}=\frac{1}{3\epsilon}\,,
\end{equation}
which agrees with the classical instanton in the quartic theory given in Eq.~(\ref{Sinst-4d}).

\subsection{Scalar Green's function on the instanton background}

In order to compute the one-loop determinant for the $\sigma$ fluctuations, we will first need to evaluate the scalar two-point function $\langle \phi^i(\vec{x}_1)\phi^j(\vec{x}_2)\rangle = \delta^{ij}G(\vec{x}_1,\vec{x}_2)$ in the instanton background. We will first perform this calculation on the $\sigma=-2$ constant solution (corresponding to moduli $\lambda=1$, $\vec{a}=0$). 

In general, for the operator 
\begin{align}
\mathcal{O}_{\vec{x}_1}= -\nabla^{2}_{\vec{x}_1}+m^{2}
\end{align}
the Green's function is given by
\begin{align}
\mathcal{O}_{\vec{x}_1}G(\vec{x}_1,\vec{x}_2)= \frac{1}{\sqrt{g}}\delta(\vec{x}_1-\vec{x}_2) \,.
\end{align}
The solution of this equation is
 \es{GGen}{
G(\vec{x}_1,\vec{x}_2)= C\cdot \,_2F_{1}(a,b,c,z) \,, 
 }
where $C$ is an $m$-dependent constant, which we will fix below. Here  $1-z=\frac{s(\vec{x}_1, \vec{x}_2)^{2}}{4}$, with $s(\vec{x}_1, \vec{x}_2)$ the chordal distance on $S^d$, and the constants $a,b,c$ are given by
\begin{align}
a=\frac{1}{2}(d-1+\sqrt{(d-1)^{2}-4m^{2}}) \,,  \quad  b=\frac{1}{2}(d-1-\sqrt{(d-1)^{2}-4m^{2}}) \,, \quad c=\frac{d}{2}\,.
\end{align}
When $m^2 = 0$, we have a conformally coupled scalar, whose Green's function $G_0(\vec{x}_1,  \vec{x}_2)$ is obtained by conformally mapping the flat space Green's function for a massless scalar, $\frac{\Gamma(\frac{d}{2}-1)}{4\pi^{d/2}} \frac{1}{\abs{\vec{x}_1 - \vec{x}_2}^{d - 2}}$:
 \es{G0}{
  G_0(\vec{x}_1,\vec{x}_2)=C_0\frac{1}{s(\vec{x}_1,\vec{x}_2)^{d - 2}} \,, \qquad
    C_0 \equiv \frac{\Gamma(\frac{d}{2}-1)}{4\pi^{d/2}}  \,.
 }
When $m^{2}= \frac{d(d-2)}{4}-2$, Eq.~\eqref{GGen} gives  
 \es{GInst}{
G(\vec{x}_1,\vec{x}_2)= C\cdot \frac{2^{d-2}}{d}\big(\frac{d-4}{s(\vec{x}_1,\vec{x}_2)^{d-2}}+ \frac{1}{s(\vec{x}_1,\vec{x}_2)^{d-4}}\big) \,.
 }
To determine the constant $C$, we note that $G(\vec{x}_1, \vec{x}_2)$ must have the same short-distance singularity as the propagator of a conformally coupled scalar, so setting equal the coefficients of $1/s(\vec{x}_1,\vec{x}_2)^{d-2}$ terms in \eqref{G0} and \eqref{GInst} we obtain $C = \frac{d\Gamma \big(\frac{d}{2}-2\big)}{2(4\pi)^{d/2}}$, and then
 \es{GInstAgain}{
G(\vec{x}_1,\vec{x}_2)= \frac{\Gamma(\frac{d}{2}-1)}{4\pi^{d/2}}\left(\frac{1}{s(\vec{x}_1,\vec{x}_2)^{d-2}}+ \frac{1}{(d-4)s(\vec{x}_1,\vec{x}_2)^{d-4}}\right) \,.
 }

The general dependence on the moduli may be reinstated by performing a conformal transformation to flat $\R^d$, where $\vec{a}$ and $\lambda$ are related simply to translations and dilations. 
One finds that for the general instanton profile, using the embedding coordinates introduced earlier, we can write
 \es{ScalarGreenGen}{
  G(\vec{x}_1, \vec{x}_2) =  \frac{\Gamma(\frac{d}{2}-1)}{4\pi^{d/2}}\left(\frac{1}{s(\vec{x}_1, \vec{x}_2)^{d-2}} +  \frac{1}{(d-4)s(\vec{x}_1, \vec{x}_2)^{d-4}} \frac{1}{(X \cdot P_1) (X \cdot P_2)} \right)\,,
 }
where $P_1$ and $P_2$ are the embedding space coordinates of the points $\vec{x}_1$ and $\vec{x}_2$.    Noting that $s(\vec{x}_1, \vec{x}_2)^2 = -2 P_1 \cdot P_2$, we can also write this formula completely in embedding space as
 \es{ScalarGreenGen2}{
  G(\vec{x}_1, \vec{x}_2) = \frac{\Gamma(\frac{d}{2}-1)}{4\pi^{d/2}}\frac{1}{(-2 P_1 \cdot P_2)^{(d-2)/2}} \left[ 1 - \frac{2P_1 \cdot P_2 }{(d-4)(X \cdot P_1) (X \cdot P_2)} \right] \,.}
This formula is true both on $\R^d$ and on $S^d$ provided we use \eqref{FlatP} or \eqref{SphereP} as appropriate.  

In the special case $d=5$, we have 
 \es{Gd5}{
  G(\vec{x}_1, \vec{x}_2) = \frac{1}{8 \pi^2} \frac{1}{s(\vec{x}_1, \vec{x}_2)^{3}} \left[  1 +  \frac{s(\vec{x}_1, \vec{x}_2)^{2}}{(X \cdot P_1) (X \cdot P_2)} \right] \,.}

\subsection{The $\sigma$ determinant}
\label{sigmadet}

As per \eqref{ZExp}, the kernel for the $\sigma$ fluctuations at any saddle is given by $-N G(\vec{x}_1, \vec{x}_2)^2$.  This kernel is diagonalized by expanding it in spherical harmonics $Y_{n, \vec{m}}(\vec{x})$, with eigenvalues $\lambda_n$:
 \es{ExpEigenvalues}{
   -N G^2(\vec{x}_1, \vec{x}_2) = \sum_{n, \vec{m} } \lambda_n  Y_{n, \vec{m}}^*(\vec{x}_1) Y_{n, \vec{m}}(\vec{x}_2) \,.
 }
Here, the $n = 0, 1, 2, \ldots$ index labels the distinct representations of the rotation group appearing in the spherical harmonic decomposition of a scalar function, while the multi-index $\vec{m}$ labels the various states in a given representation, taking $D_n$ values, with $D_n$ given in \eqref{lamDLap}.  The expansion \eqref{ExpEigenvalues} can be performed using the general formula
 \es{chord-exp}{
\frac{1}{s(\vec{x}_1,\vec{x}_2)^{2\Delta}}= \sum_{n=0}^{\infty}k_{n}(\Delta) Y_{n,\vec{m}}^{*}(\vec{x}_1)Y_{n,\vec{m}}(\vec{x}_2) \,, \qquad
 k_{n}(\Delta)= \pi^{\frac{d}{2}}2^{d-2\Delta}\frac{\Gamma(\frac{d}{2}-\Delta)\Gamma(n+\Delta)}{\Gamma(\Delta)\Gamma(d+n-\Delta)}\,.
 }

Let us start by evaluating the $\sigma$ determinant around the perturbative vacuum $\sigma=0$, where the Green's function is given in \eqref{G0}.    Applying \eqref{chord-exp} in this case, we obtain the eigenvalues
 \es{lam0}{
  \lambda^{(0)}_n = - N \left(\frac{\Gamma(\frac{d}{2}-1)}{4\pi^{d/2}}  \right)^2 k_n ( d - 2) =  \frac{N }{ 2^d  (4\pi)^{(d - 3)/2} \sin ( \frac{\pi d}{2} ) \Gamma(  \frac{d-1}{2}) }
   \frac{\Gamma( n + d - 2) }{\Gamma( n + 2) } \,.
 }
The same computation at the constant instanton saddle with the Green's function \eqref{GInst} gives
 \es{lam1}{
  \lambda^{(1)}_n =  \frac{N }{ 2^d  (4\pi)^{(d - 3)/2} \sin ( \frac{\pi d}{2} ) \Gamma(  \frac{d-1}{2}) }
   \frac{\Gamma(n + d - 2)}{\Gamma(n + 2)}  \frac{\Gamma(n + 3)}{\Gamma(n + d - 3)} 
     \frac{\Gamma(n)}{\Gamma(n + d)}  \frac{\Gamma(n + d  + 1)}{\Gamma(n - 1)}
       \frac{\Gamma(n + d - 4)}{\Gamma(n + 4)}
 }
Then
 \es{lamRatio}{
   \frac{\lambda^{(1)}_n }{\lambda^{(0)}_n }
        = \frac{(n + d) (n - 1)}{(n + d - 4) (n + 3)} \,.
 }
Note that while $\lambda^{(0)}_n$ is positive for all $n$, $\lambda^{(1)}_n$ is negative for $n=0$, it vanishes for $n=0$, and it is positive for all $n > 0$.  The negative mode at $n=0$ (note that the degeneracy is $D_0 = 1$ in this case) will give an imaginary part to the sphere free energy and other observables.  The fact that when $n=1$ we have $D_1 = d+1$ zero modes is a consequence of the existence of a whole manifold of instanton saddles \eqref{inst-Sd} parameterized by the $d+1$ parameters $\vec{a}$ and $\lambda$.  The zero modes represent infinitesimal motions along this manifold starting at the $\vec{a} = 0$ and $\lambda = 1$ saddle.  Because these saddles are related by conformal transformations, the spectrum of quadratic fluctuations in $\delta \sigma$ must be independent of the instanton moduli $\vec{a}$ and $\lambda$.

If not for the presence of zero modes, the instanton correction to the sphere free energy denoted by $ \Vol(\HH^{d+1}) A_1(d) / A_0(d)$ in \eqref{ZSd} would simply equal $\prod_n (\lambda_n^{(0)} / \lambda_n^{(1)})^{D_n/2}$.  Due to the presence of the zero modes, however, the factor $\lambda_1^{D_1/2}$ is replaced by an integral over the instanton moduli space. In addition, the divergent integral over the negative mode requires careful treatment \cite{Langer:1967ax, Callan:1977pt}. Thus, when computing the instanton correction, let us split up the $n=0$ and $n=1$ modes from the rest, and write
 \es{ARatio}{
  \Vol(\HH^{d+1}) \frac{A_1(d)}{A_0(d)} = 
   c_0 \sqrt{\abs{\frac{\lambda^{(0)}_0 }{\lambda^{(1)}_0} } }
    \left( \lambda^{(0)}_1\right)^{\frac{d + 1}{2}}  \mu_\text{zero modes}(d) R_d \,, \qquad
     R_d \equiv \prod_{n= 2}^\infty \left( \frac{\lambda^{(1)}_n }{\lambda^{(0)}_n } \right)^{-D_n / 2} \,,
 }
where $c_0$ is a factor arising from the analytically continued integral over the negative mode (to be discussed below), and the zero mode measure $\mu_\text{zero modes}(d)$ will be computed in the next subsection. The factor $R_d$ is the contribution from all the modes with $n \geq 2$, and we may compute it explicitly as follows:
 \es{GotR}{
  \log R_d = - \frac 12 \sum_{n = 2}^\infty D_{n}(d) \log \frac{(n + d) (n - 1)}{(n + d - 4) (n + 3)} \,.
 }
In $d = 5$, we obtain (see Appendix \ref{oneloopdet})
 \es{GotR5}{
  \log R_5 =  - \frac{\zeta(3)}{\pi^2} + \log \frac{9 \sqrt{15}}{2^9 \pi^3}\,.
 }
We will calculate $\mu_\text{zero modes}(d) $ in the next section. 

Let us now discuss the factor $c_0$ introduced in (\ref{ARatio}). The negative mode arises from the integration over the constant part of $\sigma$, so this is essentially a problem of analytically continuing an ordinary integral. Let us consider a concrete example \cite{Mckane:1978me} which is qualitatively similar to our situation, namely an Airy-like integral of the form
\begin{equation}
I_C = \int_C \frac{dz}{\sqrt{2\pi}} e^{-N(\frac{z^2}{2}+\frac{z^3}{3})}\,,
\end{equation}
where $C$ is a choice of contour, and we are interested in the behavior of the integral for large $N$. The function in the exponent has essentially the same form as the classical potential (\ref{V6d}) for $\sigma$ in the cubic theory near six dimensions. There are three choices of $C$ for which the integral converges for real and positive $N$: a contour $C_+$ starting at $e^{2i\pi/3}\cdot \infty$ and ending at $+\infty$; a contour $C_-$ starting at $e^{-2i\pi/3}\cdot \infty$ and ending at $+\infty$; and a contour $C_3$ starting and ending at $ e^{\mp 2i\pi/3}\cdot \infty$ (the sum $C_+-C_-+C_3$ is a closed contour). The function in the exponent in $I_C$ has saddle points at $z=0$ (the analog of the perturbative vacuum) and $z=-1$ (the analog of the instanton). The contour $C_3$ can be deformed to just pass through the $z=-1$ saddle; then $I_{C_3}$ is dominated by the instanton contribution alone and it is purely imaginary. On the other hand, the contours $C_{\pm}$ receive contributions from both saddle points and hence either one is a suitable contour for our physical application, where we want the perturbative vacuum to give the dominant contribution. Note that $C_+$ and $C_-$ can be deformed so that they run along the real axis passing through $z=0$, reaching the $z=-1$ saddle point, and then moving into the complex plane along a direction of steepest descent (either along positive or negative imaginary parts, corresponding to $C_+$ and $C_-$ respectively). Hence, the imaginary part of the integral at large $N$ receives contribution from the semi-infinite arc of such contours starting at $z=-1$. In the saddle point approximation this gives half of a Gaussian integral, so that
\begin{equation}
{\rm Im}(I_{C_{\pm}}) \approx \mp \frac{i}{2} \frac{e^{-N/6}}{\sqrt{N}}\,.
\end{equation}
The potential for $\sigma$ in the large $N$ theory is a more complicated function, but it is qualitatively similar.  See Figure~\ref{fig:potS5}. So we expect a similar analysis to go through in this case as well (see \cite{Callan:1977pt} for a general discussion). In particular we expect the appropriate contour to run along the real axis passing through $\sigma=0$ and reaching $\sigma=-2$, and then to move into the complex plane into either positive or negative directions of the imaginary axis. Therefore, for the purpose of extracting the imaginary parts of physical quantities due to the instanton, we should set in (\ref{ARatio})
 \es{c0Choice}{
c_0 =\pm \frac{i}{2}\,.
 }
The choice of sign, corresponding to the ambiguity in the choice of contour, gives two distinct ``complex CFTs" that are related to each another by complex conjugation.

\subsection{Measure on the instanton moduli space}

First, let us work around $\rho = 0$ ($\lambda=1$) in the instanton moduli space.  For small fluctuations, we can expand the fluctuation $\delta \sigma$ in terms of modes $\psi_a$
 \es{deltasig}{
  \delta \sigma = \sum_a c_a \psi_a
 }
that we can unit normalize
 \es{NormModes}{
  \int_{S^5} d^5 x \sqrt{g} \,  \psi_a^* \psi_b = \delta_{ab}  \,.
 }
Then, the path integral measure is
 \es{PatIntMeasure}{
  \prod_a \frac{dc_a}{\sqrt{2 \pi}} \,.
 }

We separate these modes into zero modes indexed by $A = 1, \ldots, d+1$ and non-zero modes indexed by $i$:  $c_a = (c_A, c_i)$.  If the coordinates $v^A$ parameterize the instanton moduli space, then
 \es{psiA}{
  \psi_A = \frac{1}{\sqrt{N_A}} \frac{d \sigma}{d v^A} \,,
 } 
where $N_A$ is the norm of $d\sigma / dv^A$.  From $\psi_A dc_A = (d \sigma / dv^A) dv^A$, we then conclude that $dc_A = \sqrt{N_A} dv_A$, and so the zero mode measure is
 \es{zeroModeMeasure}{
  \prod_A \frac{dc_A}{\sqrt{2 \pi}} = \prod_A dv_A \sqrt{\frac{N_A}{2 \pi}} \,.
 }

Let us now compute $N_A$.   Close to $\rho = 0$, the instanton profile is approximately $\sigma \approx - 2 - 4 \hat p \cdot (\hat n \rho) + \cdots$.  We can thus take $v_A = \rho n_A$ close to $\rho = 0$, so 
 \es{dsdv}{
  \frac{d \sigma}{d v_A} = 4 \hat p_A  \qquad \Longrightarrow \qquad
   N_A = \frac{16 \Vol(S^d) }{d+1} \,.
 }
So the zero mode measure close to $\rho = 0$ is 
  \es{MeasureOrigin}{
   \left(  \frac{16 \Vol(S^d)}{2 \pi( d+1) } \right)^\frac{d+1}{2}  \prod_A dv_A
    = \left(  \frac{16 \Vol(S^d)}{2 \pi( d+1) } \right)^\frac{d+1}{2}  \, \rho^d d \rho \vol_{S^d} \,.
   }
The quantity $ \rho^d d \rho \vol_{S^d}$ is the small $\rho$ approximation to the volume of hyperbolic space of unit radius.  Thus, the  full zero mode measure is
 \es{MeasureFull}{
  \mu^\text{zero modes}_d = \Vol(\HH^{d+1}) \left(  \frac{16 \Vol(S^d)}{2 \pi( d+1) } \right)^\frac{d+1}{2} \,.
 }
For $d = 5$, we have $\Vol(S^5) = \pi^3$, so we obtain 
 \es{d5Measure}{
  \mu^\text{zero modes}_5 = \Vol(\HH^6) \left(  \frac{4\pi^2 }{3 } \right)^3 \,.
  }

\subsection{Sphere free energy in $d=5$}

Let us finally compute the instanton contribution $\Vol(\HH^6 ) A_1 / A_0\,  e^{-N(f_1 - f_0)}$ to the sphere free energy in $d=5$.  We have the contribution of the $n=0$ modes:
 \es{nZero}{
  - \frac 12 \log \abs{\frac{\lambda_0^{(1)}}{\lambda_0^{(0)}}}  = \frac 12 \log \frac 35 \,;
 }
the contribution of the $n=1$ modes:
 \es{n1Modes}{
  \log \mu^\text{zero modes}_d  + \frac{d+1}{2} \log \lambda_1^{(0)} = \log \frac{N^3 \pi^3 \Vol(\HH^6)}{2^{15}} \,;
 } 
and, lastly, we have the contribution of the modes with $n \geq 2$ in \eqref{GotR5}:
  \es{Nonzero}{
  \log R_5 =  - \frac{\zeta(3)}{\pi^2} + \log \frac{9 \sqrt{15}}{2^9 \pi^3}   \,.
 }

Adding these expression together, exponentiating, and multiplying by $e^{-N (f_1 - f_0)}$ we find 
 \es{InstContribNoVol}{
   \frac{A_1}{A_0} e^{-N(f_1 - f_0)} = c_0   \left( \frac{3 N}{256} \right)^3  \exp \left[ - N \left( \frac{1}{8} \log 2 - \frac{3 \zeta(3)}{16\pi^2} \right)  - \frac{\zeta(3)}{\pi^2}  \right] \,.
 }
The expression that appears in the instanton contribution to the $S^5$ free energy in \eqref{ZSd} also includes a factor of the volume of the unit radius hyperbolic space, $\Vol(\HH^6)$.  For general $d$, one can regularize such a factor using a regulator preserving spherical symmetry.  The regularized value is
\begin{equation}
\Vol(\HH^{d+1}) = \pi^{d/2} \Gamma \left(-\frac{d}{2}\right) \,.
\end{equation}
Thus, the regularized value of $\Vol(\HH^6)$ in \eqref{ZSd} should be taken to be $\Vol(\HH^6) = -8 \pi^3 / 15$.  Putting everything together, the imaginary part of the $S^5$ free energy is then approximately given by 
\begin{equation}
{\rm Im}(F_{S^5}) \approx -  \Vol(\HH^6) \frac{A_1}{A_0} e^{-N(f_1 - f_0)}  =  \pm \frac{1}{30} \left( \frac{3 \pi N}{128} \right)^3  \exp \left[ - N \left( \frac{1}{8} \log 2 - \frac{3 \zeta(3)}{16\pi^2} \right)  - \frac{\zeta(3)}{\pi^2}  \right] \,,
\end{equation}
where we have used $c_0=\pm i/2$ as explained in Section~\ref{sigmadet}. 

Note that while an overall factor of $\Vol(\HH^6)$ arises in the instanton contribution to the sphere free energy, such a factor will be absent in the computations of other CFT observables presented below (which, in particular, are independent of whether we choose spherical or planar slicing for the hyperbolic space metric).

\section{Imaginary parts of scaling dimensions}
\label{SCALING}

A key property of complex CFTs is the presence of some operators whose scaling dimensions are complex. This behaviour has been found in a variety of models,
including \cite{Dymarsky:2005uh,Pomoni:2008de,Kaplan:2009kr,
Giombi:2015haa,Grabner:2017pgm,Gorbenko:2018ncu,Gorbenko:2018dtm,Benini:2019dfy}. We will now explore how the instanton effects contribute small imaginary parts to the operator scaling dimensions. While in a number of large $N$ theories the complex scaling dimensions have been found to lie on the principal series $d/2 + i \alpha$, we find that the instanton effects produce complex dimensions of a more general form, with real parts not necessarily equal to $d/2$. 

\subsection{Anomalous dimension of $\phi_i$}

For the perturbative saddle, at leading order in $1/N$, the two point function of $\phi_i$ is
 \es{phiTwoLeading}{
  \langle \phi_i(\vec{x}_1) \phi_j(\vec{x}_2) \rangle_0 = \delta_{ij} \frac{\Gamma( \frac d2 - 1)}{4 \pi^{\frac d2}} \frac{1}{s(\vec{x}_1, \vec{x}_2)^{d-2}} 
   = \delta_{ij} \frac{1}{8 \pi^2} \frac{1}{s(\vec{x}_1,\vec{x}_2)^3} \,,
 }
where, as before, $s(\vec{x}_1,\vec{x}_2)$ is either $\abs{\vec{x}_1 - \vec{x}_2}$ on $\R^5$ or the chordal distance on $S^5$.  The exponent in the denominator shows that the scaling dimension of $\phi_i$ equals $3/2$ at large $N$.   The expression \eqref{phiTwoLeading} receives both perturbative and non-perturbative corrections in $1/N$.  While the perturbative corrections are real, the first contribution to the imaginary part comes from the instanton saddles discussed in the previous section.

Let us denote the scaling dimension of $\phi_i$ by $\Delta_\phi = 3/2 + \delta_\phi$.  Since the instanton contribution to $\delta_\phi$, which we denote by $\delta_\phi^\text{inst}$, is small, its effect is to modify the exponent in \eqref{phiTwoLeading} only slightly.  Expanding $1/s(\vec{x}_1, \vec{x}_2)^{3 + 2 \delta} \approx (1 - 2 \delta \log [ s(\vec{x}_1, \vec{x}_2) / \epsilon] )  /s(\vec{x}_1, \vec{x}_2)^{3}$, where $\epsilon$ is the UV cutoff, we find that the instanton contribution to the $\phi_i$ two-point function must take the form
 \es{phiFull}{
  \langle \phi_i(\vec{x}_1) \phi_j(\vec{x}_2) \rangle_\text{inst}
   = -\delta_{ij} \frac{1}{4 \pi^2} \frac{\log [s(\vec{x}_1, \vec{x}_2) / \epsilon ] }{s(\vec{x}_1, \vec{x}_2)^{3}} \delta_\phi^\text{inst} \,.
 }
To obtain \eqref{phiFull}, note that when computing the two-point function $ \langle \phi_i(\vec{x}_1) \phi_j(\vec{x}_2) \rangle$ we should perform the path integral with a $\phi_i(\vec{x}_1) \phi_j(\vec{x}_2)$ insertion by summing over all saddle points, and then divide the answer by the partition function.  Thus
 \es{phiTot}{
  \langle \phi_i(\vec{x}_1) \phi_j(\vec{x}_2) \rangle
   = \delta_{ij} \frac{A_0 e^{-N f_0}  G_0 (\vec{x}_1, \vec{x}_2)+ A_1 e^{-N f_1}   \int_{\HH^6} dX \, G(\vec{x}_1, \vec{x}_2)  + \cdots }{A_0 e^{-N f_0} +  \Vol(\HH^6) A_1 e^{-N f_1} + \cdots} \,,
 }
where for each saddle point we only included the contribution from the $\phi_i$ and $\sigma$ determinants.  Expanding \eqref{phiTot} to first order in $A_1$, Eq.~\eqref{phiTot} equals
 \es{phiTotExpansion}{
   \delta_{ij} G_0(\vec{x}_1, \vec{x}_2)+ \delta_{ij} \frac{A_1}{A_0} e^{-N (f_1 - f_0)} \left[- \Vol(\HH^6) G_0(\vec{x}_1, \vec{x}_2) +  \int_{\HH^6} dX \, G(\vec{x}_1, \vec{x}_2) \right] + \cdots \,.
 }
 Using \eqref{phiTwoLeading} and \eqref{Gd5}, we then identify
 \es{phiNSAgain}{
  \langle \phi_i(\vec{x}_1) \phi_j(\vec{x}_2) \rangle_\text{inst}
   =  \frac{\delta_{ij} }{8 \pi^2}  \frac{A_1}{A_0} e^{-N(f_1 - f_0)} 
     \frac{I_\phi(\vec{x}_1,\vec{x}_2)}{s(\vec{x}_1, \vec{x}_2)^{3}}   \,, \qquad
      I_\phi(\vec{x}_1,\vec{x}_2) \equiv \int_{\HH^6} dX \,  \frac{-2 P_1 \cdot P_2}{(X \cdot P_1) (X \cdot P_2)} \,.
 }
Comparing this expression with \eqref{phiFull}, we conclude
 \es{Gotdelta}{
  \delta_\phi^\text{inst} =  
   \frac{ 1}{2}  \frac{A_1}{A_0} 
    e^{-N(f_1 - f_0)}  \frac{dI_\phi(\vec{x}_1,\vec{x}_2)}{d\log \epsilon}   
       \,.
 } 
To calculate $\delta_\phi^\text{inst}$, all that remains to do is to calculate the logarithmic derivative of $I_\phi$ with respect to $\epsilon$.  Note that $I_\phi(\vec{x}_1,\vec{x}_2) $ is Weyl-invariant, so it is independent on whether the theory is placed on $\R^5$ or on $S^5$.  

\subsubsection{Hard cutoff regularization}

Since the logarithmic derivative should not depend on the points $\vec{x}_1 $ and $\vec{x}_2$, let us set $\vec{x}_1 = 0$ and $\abs{\vec{x}_2} \to \infty$.  On $S^5$ these two points correspond to the North and South poles of $S^5$, respectively.  In embedding space, the North pole corresponds to $P_N = (\vec{0}, 1, 1)$ and the South pole corresponds to $P_S = (\vec{0}, -1, 1)$, so the chordal distance between them is $s(N, S) = 2$. Then, taking $X = (\hat n_4 \sin \theta \sinh \rho ,  \cos \theta \sinh \rho, \cosh \rho)$, where $\hat n_4$ is a unit vector on $S^4$, we have
 \es{Integral}{
  I_\phi(N, S) = \int_{\HH^6} dX \,  \frac{s(N, S)^{2}}{(X \cdot P_N) (X \cdot P_S)} 
   = \Vol(S^4) \int d\rho\, d\theta \, \sinh^5 \rho \sin^4 \theta
    \frac{4}{1 +  \sinh^2 \rho \sin^2 \theta} \,.
 }
(Here, we used the $SO(5)$ rotational symmetry of this configuration to pull out a volume factor of $S^4$.)   
Both integrals in \eqref{Integral} can be done analytically, but for the $\rho$ integral we impose an upper cutoff:
  \es{Integral2}{
   I_\phi(N, S) = 2 \pi  \Vol(S^4) \int_0^{\rho_m} d\rho\, \sinh^5 \rho\, \frac{\cosh \rho + 2}{\cosh \rho (\cosh \rho + 1)^2} 
    = \frac{2 \pi^3 e^{3 \rho_m}}{9}
     - \frac{22 \pi^3 e^{\rho_m}}{3} + \frac{32 \pi^3}{3}  \rho_m + \cdots \,,
  }
where we used $\Vol(S^4) = 8 \pi^2/3$.  Taking $e^{\rho_m} = 1/\epsilon$ (if we were to reinstate the radius $R$ of $S^5$ we would have $e^{\rho_m} = R/\epsilon$), we then find 
 \es{Integral3}{
  I_\phi(N, S) = \text{(power-law divergence)} - \frac{32 \pi^3}{3}  \log \epsilon + \cdots \,.
 }
From this expression we can extract the logarithmic derivative 
 \es{logDer}{
  \frac{d I_\phi(\vec{x}_1, \vec{x}_2)}{d \log \epsilon}  =- \frac{32 \pi^3}{3} \,, 
 }
where we used the fact that the logarithmic derivative should be independent of the choice of points.

\subsubsection{Analytic regulator in Poincar\'e coordinates}

Let us now obtain the same result using a different regulator.  We could also write $I_\phi(\vec{x}_1, \vec{x}_2)$ in Poincar\'e coordinates:
  \es{int}{
I_\phi(\vec{x}_1,\vec{x}_2) 
  = 4 \abs{\vec{x}_1  - \vec{x}_2}^2 \int \frac{dz d^5 \vec{a}}{z^6} \frac{z^2}{\left[ z^2 + (\vec{x}_1 - \vec{a})^2 \right]  \left[ z^2 + (\vec{x}_2 - \vec{a})^2\right]} \,.
 } 
If we regularize this integral by multiplying the integrand by $(z/\epsilon)^{s}$, we can use the formulas from \cite{Freedman:1998tz} to evaluate \eqref{int} to
 \es{intFreedman}{
  I_\phi(\vec{x}_1,\vec{x}_2) = \frac{32 \pi^3}{3 s} - \frac{32 \pi^3}{3} \log (\epsilon / \abs{\vec{x}_1 - \vec{x}_2}) + \cdots \,.
 }
The logarithmic derivative of this expression with respect to $\epsilon$ reproduces \eqref{logDer}.
 
\subsection{Anomalous dimension of $\sigma$} 
 
We can perform a similar calculation to determine the leading contribution to the imaginary part of the scaling dimension of $\sigma$.  The leading two-point function of $\sigma$ can be found from the perturbative saddle by inverting the kernel $-N G_0(\vec{x}_1, \vec{x}_2)^2$ multiplying the $\sigma$ fluctuations.  This is easily done after performing the spherical harmonic decomposition as in \eqref{chord-exp}, whereby this kernel becomes $-N C_0^2 k_n(d-2)$, with $C_0$ being the constant defined in \eqref{G0}.  Because the inverse of this kernel gives $ -1/( N C_0^2 k_n(d-2)) = C_\sigma k_n(2)$, with 
 \es{GotCsigma}{
  C_\sigma = -\frac{1}{N C_0^2 k_n(2) k_n(d-2)}  = \frac{8 (d-4) \Gamma(d-2) \sin \frac{\pi d}{2}}{N \pi \Gamma(\frac d2 - 1)^2 }
   = \frac{64}{\pi^2 N} \,, 
 }
one finds that the leading order two-point function of $\sigma$ is
 \es{sigmaTwoPoint}{
  \langle \sigma (\vec{x}_1) \sigma (\vec{x}_2) \rangle_0 = \frac{C_\sigma}{s(\vec{x}_1, \vec{x}_2)^4} \,.
 }
As in the case of $\phi$, the dimension of $\sigma$  is $\Delta_\sigma = 2 + \delta_\sigma$, so the contribution of the instanton saddle to the two-point function must take the form
 \es{sigmaFull}{
  \langle \sigma(\vec{x}_1) \sigma(\vec{x}_2) \rangle_\text{inst}
   = - 2 C_\sigma \frac{\log [s(\vec{x}_1, \vec{x}_2) / \epsilon ] }{s(\vec{x}_1, \vec{x}_2)^{4}} \delta_\sigma^\text{inst} \,,
 }
where $\epsilon$ is the UV cutoff. 

To leading order in $N$, the contribution from the instanton saddle can be found by simply replacing $\sigma$ by its classical value $\sigma = -8 / (-2 X \cdot P)^2$ and computing the functional integral over $\sigma$ in the presence of the insertion $\sigma^2$.  This integral yields the same answer as without the insertion for the non-zero modes.  The integration over the instanton moduli space now replaces $\Vol(\HH^6)$ in $\mu_d^\text{zero modes}$ with the integral of $\sigma^2$ over $\HH^6$. Thus,
 \es{sigmaTot}{
  \langle \sigma(\vec{x}_1) \sigma(\vec{x}_2) \rangle
   =  \frac{A_0 e^{-N f_0}  \langle \sigma (\vec{x}_1) \sigma (\vec{x}_2) \rangle_0 + A_1 e^{-N f_1}   \int_{\HH^6} dX \, \frac{ -8}{  (-2 X \cdot P_1)^2 } 
    \frac{ -8}{  (-2 X \cdot P_2)^2 } + \cdots }{A_0 e^{-N f_0} + \Vol(\HH^6) A_1 e^{-N f_1} + \cdots} \,,
 }
where we included only the leading contributions from the two saddles.  Expanding at small $A_1$, we identify the instanton contribution to the $\sigma$ two-point function to be 
 \es{InstSaddlesigma}{
  \langle \sigma(\vec{x}_1) \sigma(\vec{x}_2) \rangle_\text{inst}
   =   \frac{A_1}{A_0} e^{-N(f_1 - f_0)} 
    \int_{\HH^6} dX \,  \frac{4 }{(X \cdot P_1)^2(X \cdot P_2)^2}  
    \,.
 }
Note that the expansion of the denominator in \eqref{sigmaTot} generates a term proportional to $\langle \sigma (\vec{x}_1) \sigma (\vec{x}_2) \rangle_0$ that is proportional to $A_1 / A_0$, but this term is suppressed by one power of $1/N$ relative to the term written in \eqref{InstSaddlesigma}. Comparing this expression with \eqref{sigmaFull}, we extract
 \es{Gotdeltasigma}{
  \delta_\sigma^\text{inst} = \frac{2}{ C_\sigma}  \frac{A_1}{A_0} e^{-N(f_1 - f_0)} 
    \frac{d  I_\sigma(\vec{x}_1, \vec{x}_2)}{d \log \epsilon}   \,, \qquad
    I_\sigma(\vec{x}_1, \vec{x}_2) \equiv  \int_{\HH^6} dX \,  \frac{ (- 2 P_1 \cdot P_2)^2 }{(X \cdot P_1)^2 (X \cdot P_2)^2} \,.
 }
 Like $I_\phi(\vec{x}_1,\vec{x}_2) $, the quantity $I_\sigma(\vec{x}_1,\vec{x}_2) $ is also Weyl-invariant, so it is independent on whether the theory is placed on $\R^5$ or on $S^5$.

\subsubsection{Hard cutoff regularization}

Let us now compute the logarithmic derivative of $I_\sigma(\vec{x}_1, \vec{x}_2)$ with respect to teh UV cutoff $\epsilon$ by using a hard cutoff. Taking $\vec{x}_1 = N$ and $\vec{x}_2 = S$, we have
 \es{IntegralSigma}{
  I_\sigma(N, S) = \int_{\HH^6} dX \,  \frac{s(N, S)^{4}}{(X \cdot P_N)^2 (X \cdot P_S)^2} 
   = \Vol(S^4) \int d\rho\, d\theta \, \sinh^5 \rho \sin^4 \theta
    \frac{16}{\left( 1 +  \sinh^2 \rho \sin^2 \theta\right)^2} \,.
 }
We can again do both integrals:
  \es{Integral2sigma}{
   I_\sigma(N, S) = 8 \pi  \Vol(S^4) \int_0^{\rho_m} d\rho\, \sinh^5 \rho\, \frac{2 \cosh \rho + 1}{\cosh^3 \rho (\cosh \rho + 1)^2} 
    = 
     \frac{64 \pi^3 e^{\rho_m}}{3} - 64 \pi^3  \rho_m + \cdots \,.
  }
Using $e^{\rho_m} = 1/\epsilon$, we obtain
 \es{Integral3sigma}{
  I_\sigma(N, S) = \text{(power-law divergence)} + 64 \pi^3  \log \epsilon + \cdots \,.
 }
Making use of the fact that the logarithmic derivative of $I_\sigma$ is independent of the choice of points, we conclude that 
 \es{IsigLogDer}{
   \frac{d I_\sigma(\vec{x}_1, \vec{x}_2)}{\log \epsilon}  =  64 \pi^3  \,.
 }

\subsubsection{Analytic regulator in Poincar\'e coordinates}

We can also obtain the same result using an analytic regulator.  In Poincar\'e coordinates, we have
  \es{intsigma}{
I_\sigma(\vec{x}_1,\vec{x}_2) 
  = 16 \abs{\vec{x}_1  - \vec{x}_2}^4 \int \frac{dz d^5 \vec{a}}{z^6} \frac{z^4}{\left[ z^2 + (\vec{x}_1 - \vec{a})^2 \right]^2  \left[ z^2 + (\vec{x}_2 - \vec{a})^2\right]^2} \,.
 } 
Let's again regularize the integral by multiplying the integrand by $(z/\epsilon)^{s}$:
 \es{intFreedmansigma}{
  I_\sigma(\vec{x}_1,\vec{x}_2) = -\frac{64 \pi^3}{s} + 64 \pi^3 \log (\epsilon / \abs{\vec{x}_1 - \vec{x}_2}) + \cdots \,.
 }
A derivative of \eqref{intFreedmansigma} with respect to $\log \epsilon$ then reproduces \eqref{IsigLogDer}.

\subsection{Numerical values}

Since the leading imaginary part of the scaling dimensions comes from the instanton contribution, we have
 \es{Im}{
  \Im \Delta_\phi &\approx \mp \frac{16 \pi^3}{3} \abs{\frac{A_1}{A_0} e^{-N (f_1 - f_0)}} \,, \\
  \Im \Delta_\sigma &\approx  \pm 2 \pi^5 N \abs{\frac{A_1}{A_0} e^{-N (f_1 - f_0)}} \,,   
 }
where the quantity $\frac{A_1}{A_0} e^{-N (f_1 - f_0)}$ is given explicitly in \eqref{InstContribNoVol}, and the overall sign choice corresponds to $c_0 = \pm i/2$ as in \eqref{c0Choice}.  

In $d=5$, if we require $\abs{\Im \Delta_\phi} < 10^{-2}$ or $10^{-3}$, then we find $N > 172$ or $N>220$, respectively.
If we require $\abs{\Im \Delta_\sigma} < 10^{-2}$ or $10^{-3}$, we find $N > 310$ or $N > 355$, respectively. These constraints are roughly commensurate with the smallest values of $N$ where the ``islands" in the bootstrap for the $d=5$ $O(N)$ model were observed \cite{Li:2016wdp}.\footnote{While the values of $N$ explored in \cite{Li:2016wdp} extended up to $500$, it would be desirable to explore even higher values (for example, around $1000$), both because the imaginary parts of scaling dimensions are much smaller there, and because the $1/N$ expansion is more reliable.} 

\section{Instanton contribution to other quantities}
\label{COEFFICIENTS}

 \subsection{$O(N)$ current two-point function coefficient $c_J$}

The scaling dimensions are not the only quantities that acquire imaginary contributions from the instanton background.  Another quantity is the coefficient $c_J$ that appears in the two-point function of the canonically-normalized $O(N)$ current $j_{\mu ij}$.  In flat space, the position dependence of this two-point function is fixed by the conformal symmetry, and the overall normalization defines $c_J$:
 \es{TwoPointCurrent}{
  \langle j_{\mu ij}(\vec{x}_1)j_{\nu kl} (\vec{x}_2)  \rangle
   &=  c_J \frac{2 C_0^2 (d-2)}{\abs{\vec{x}_1 - \vec{x}_2}^{2d -2}} I_{\mu\nu} (\delta_{ik} \delta_{jl} - \delta_{il} \delta_{jk} )\,, \\
   I_{\mu\nu} &\equiv  \delta_{\mu\nu}  - 2 \frac{(\vec{x}_1 - \vec{x}_2)_\mu (\vec{x}_1 - \vec{x}_2)_\nu}{\abs{\vec{x}_1 - \vec{x}_2}^2}  \,,
 } 
where $C_0$ is the constant defined in \eqref{G0}.  This definition of $c_J$ is such that a free theory of $N$ massless scalar fields, which has an $O(N)$ global symmetry under which the scalar fields transform as a vector, has $c_J = 1$.

For us (or for a theory of $N$ free massless scalars $\phi_i$), the canonically-normalized $O(N)$ current is
 \es{ONCurrent}{
  j_{\mu ij} =  \phi_i \partial_\mu \phi_j - \phi_j \partial_\mu \phi_i  \,.
 }
Canonical normalization means that the leading term in the OPE between $j_{\mu ij}$ and an operator ${\cal O}_i$ transforming in the vector representation of $O(N)$ takes the form
 \es{CanNorm}{
  j^{\mu}_{ij}(\vec{x}) {\cal O}_k (0) = -\frac{1}{(d-2) \Vol(S^{d-1}) }\frac{x^\mu}{\abs{\vec{x}}^d} \big(\delta_{ik} {\cal O}_j (0)- \delta_{jk} {\cal O}_i(0) \big)+ \text{less singular terms} \,,
 }
as $\abs{\vec{x}} \to 0$.  This equation is such that if we construct the charge operator $Q^\Sigma_{ij} = \int_\Sigma j^\mu_{ij} n_\mu$ associated with a closed surface $\Sigma$ with outward pointing normal $n_\mu$ surrounding the origin, then $Q^\Sigma_{ij}$ acts on ${\cal O}_k(0)$ as a generator of the $O(N)$ symmetry:
 \es{QO}{
  Q^\Sigma_{ij} {\cal O}_k(0) = \delta_{ik} {\cal O}_j(0) - \delta_{jk} {\cal O}_i(0) \,.
 }
The equation \eqref{QO} is true provided that $\Sigma$ does not enclose any other $O(N)$-non-invariant operator other than ${\cal O}_k(0)$.  By setting ${\cal O}_k = \phi_k$, we can immediately check that the current defined in \eqref{ONCurrent} obeys \eqref{CanNorm} at leading order in large $N$ for both the perturbative saddle and the instanton saddles.  This property follows from the fact that the short distance singularity of the Green's function is precisely the same for both saddles.

Working around either the perturbative saddle or one of the instanton saddles, we can write
 \es{jjCorr}{
  \langle j_{\mu ij}(\vec{x}_1)j_{\nu kl} (\vec{x}_2)  \rangle
   = 2 \biggl[ G \partial_\mu^{(1)}  \partial_\nu^{(2)} G - \partial_\mu^{(1)} G \partial_\nu^{(2)} G \biggr] 
    (\delta_{ik} \delta_{jl} - \delta_{il} \delta_{jk} ) \,.
 }
It is straightforward to check that with $G = G_0$ given in \eqref{G0}, one reproduces \eqref{TwoPointCurrent} with $c_J = 1$.  After using \eqref{ScalarGreenGen} or \eqref{ScalarGreenGen2} around a given instanton saddle specified by the moduli $(\lambda, \vec{a})$ and integrating over these moduli, Eq.~\eqref{jjCorr} becomes
 \es{jjinst}{
   \langle j_{\mu ij}(\vec{x}_1)j_{\nu kl} (\vec{x}_2)  \rangle_\text{inst $(\lambda, \vec{a})$}
    &=  C_0^2 (\delta_{ik} \delta_{jl} - \delta_{il} \delta_{jk} ) 
     \biggl[ 
       \frac{I_{\mu\nu}}{\abs{\vec{x}_1 - \vec{x}_2}^{2d -2}} 
        \biggl(2 (d-2) 
         + \frac{16 (d-2)}{d - 4}  Q
          + \frac{32}{d-4} Q^2 \biggr)  \\
      &{}+ \frac{8}{d-4} \frac{1}{\abs{\vec{x}_1 - \vec{x}_2}^{2d -4}} \partial_\mu^{(1)} \partial_\nu^{(2)} Q        
     \biggr]   \,,
 }
with the quantity $Q$ defined as
 \es{QDef}{
  Q \equiv \frac{\lambda^2 \abs{\vec{x}_1 - \vec{x}_2}^2}{( 1+ \lambda^2 (\vec{x}_1 - \vec{a})^2)
   ( 1+ \lambda^2 (\vec{x}_2 - \vec{a})^2)} 
    = \frac{-P_1 \cdot P_2}{2 (P_1 \cdot X) (P_2 \cdot X)} \,.
 }
From \eqref{jjinst} one can obtain the full instanton contribution (up to one-loop order) after performing an integral over the instanton moduli space:
 \es{InstFull}{
  \langle j_{\mu ij}(\vec{x}_1)j_{\nu kl} (\vec{x}_2)  \rangle_\text{inst}
   &= \frac{A_1}{A_0 } e^{-N (f_1 - f_0)} \biggl[  \int_{\HH^{d+1}}  dX \, \langle j_{\mu ij}(\vec{x}_1)j_{\nu kl} (\vec{x}_2)  \rangle_\text{inst $(\lambda, \vec{a})$}  \\
    &{}- \Vol(\HH^{d+1}) \langle j_{\mu ij}(\vec{x}_1)j_{\nu kl} (\vec{x}_2)  \rangle_0 \biggr]\,,
 } 
where the term in the second line arises precisely in the same way that the $G_0$ term in the square bracket of \eqref{phiTotExpansion}.

Let us now work in $d=5$, where 
 \es{jjInst}{
  \langle j_{\mu ij}(\vec{x}_1)j_{\nu kl} (\vec{x}_2)  \rangle_\text{inst} &= 
    \frac{A_1}{A_0 } e^{-N (f_1 - f_0)}  C_0^2 (\delta_{ik} \delta_{jl} - \delta_{il} \delta_{jk} )  \\
   &{}\times  
     \int_{\HH^6} dX\, \biggl[ 
       \frac{I_{\mu\nu} (48  Q
          + 32 Q^2) }{\abs{\vec{x}_1 - \vec{x}_2}^8}   
      +  \frac{8}{\abs{\vec{x}_1 - \vec{x}_2}^6} \partial_\mu^{(1)} \partial_\nu^{(2)} Q        
     \biggr]  \,.
 } 
The integral over $X$ can be performed as in the previous section.  After regularization, we have
 \es{FirstInt}{
  \int_{\HH^6}  d X\, \left(
          48Q
          + 32 Q^2 \right)
          = \frac{320 \pi^3}{3} \,,
 } 
 and this integral is independent of weather we use the rotationally-invariant cutoff regulator or the analytic one.  From \eqref{intFreedman}, we also have
  \es{SecondInt}{
    \frac{1}{\abs{\vec{x}_1 - \vec{x}_2}^6} \partial_\mu^{(1)} \partial_\nu^{(2)} \int_{\HH^6}  d X\, 8 Q 
     = - \frac{64 \pi^3}{3} \frac{I_{\mu\nu}}{\abs{\vec{x}_1 - \vec{x}_2}^8}  \,,
  }
which again is independent of which regulator we use.  Plugging \eqref{FirstInt} and \eqref{SecondInt} into \eqref{jjInst} and comparing with the definition of $c_J$ in \eqref{TwoPointCurrent}, we find that the leading approximation for the imaginary part of $c_J$ is
 \es{cJInst}{
  \Im c_J \approx \abs{c_{J, \text{inst}}} \approx  \pm  \frac{128 \pi^3}{9}   \abs{\frac{A_1}{A_0 } e^{-N (f_1 - f_0)}}  \,,
 }
with $\frac{A_1}{A_0} e^{-N (f_1 - f_0)}$ given in \eqref{InstContribNoVol}.

\subsection{Three-point function coefficients}

Another example of a quantity that acquires an imaginary part due to the instanton contribution is the three-point function $\langle \sigma \sigma \sigma \rangle$.  Conformal invariance implies
 \es{ThreePoint}{
  \langle \sigma(\vec{x}_1) \sigma(\vec{x}_2) \sigma(\vec{x}_3) \rangle 
   = \frac{C_{\sigma \sigma \sigma}}{\abs{\vec{x}_1 - \vec{x}_2}^{\Delta_\sigma} 
     \abs{\vec{x}_1 - \vec{x}_3}^{\Delta_\sigma}
     \abs{\vec{x}_2 - \vec{x}_3}^{\Delta_\sigma}} \,,
 }
for some numerical coefficient $C_{\sigma \sigma \sigma}$.   The perturbative contribution to $C_{\sigma \sigma \sigma}$ is $O(1/N^2)$, and since it is real, we are not concerned with it here.  The contribution from the instanton saddle gives the leading imaginary part of $C_{\sigma \sigma \sigma}$, and can be evaluated simply by plugging in the classical value of $\sigma$ and integrating over the instanton moduli space.  By analogy with the expression \eqref{InstSaddlesigma} for the two-point function, we have
 \es{InstSaddlesigmaThree}{
  \langle \sigma(\vec{x}_1) \sigma(\vec{x}_2) \sigma(\vec{x}_3)  \rangle_\text{inst}
   =   \frac{A_1}{A_0} e^{-N(f_1 - f_0)} 
    \int_{\HH^6} dX \,  \frac{-8 }{(X \cdot P_1)^2(X \cdot P_2)^2 (X \cdot P_3)^2}  
    \,.
 }
This integral is convergent and was evaluated in \cite{Freedman:1998tz}:
 \es{intAdSThree}{
  \int_{\HH^6} dX\, \frac{1}{(-2 X \cdot P_1)^2 (-2 X \cdot P_2)^2 (-2 X \cdot P_3)^2 } = \frac{\pi^3}{2}
    \frac{1}{\abs{\vec{x}_1 - \vec{x}_2}^{2} 
     \abs{\vec{x}_1 - \vec{x}_3}^{2}
     \abs{\vec{x}_2 - \vec{x}_3}^{2}} \,.
 }
It follows that the leading imaginary contribution to $C_{\sigma \sigma \sigma}$ is
 \es{ImCsss}{
  \Im C_{\sigma \sigma \sigma} \approx \mp \abs{C_{\sigma \sigma \sigma, \text{inst}}}  \approx  \mp 256 \pi^3 \abs{\frac{A_1}{A_0} e^{-N(f_1 - f_0)}} \,, 
 }
with $\frac{A_1}{A_0} e^{-N (f_1 - f_0)}$ given in \eqref{InstContribNoVol}.

While the three-point function $\langle \sigma \sigma \sigma \rangle$ and consequently $C_{\sigma \sigma \sigma}$ depends on the normalization of the operator $\sigma$, one can define the normalization-independent ratio 
 \es{Ratio}{
  r = \frac{C_{\sigma \sigma \sigma}}{\langle \sigma(\hat{e}) \sigma(0) \rangle^{3/2}} \,,
 }
where $\hat e$ is a unit vector.  The leading order imaginary part of $r$ is then
 \es{leadingr}{
  \Im r \approx  \frac{\Im C_{\sigma \sigma \sigma}}{C_\sigma^{3/2}} \approx \mp \frac{\pi^6}{2} N^{3/2}  \abs{\frac{A_1}{A_0} e^{-N(f_1 - f_0)}}  \,.
 }

\section{Saddle points with $k>1$}
\label{OTHERSADDLES}

As shown in Section~\ref{largeN-inst}, the large $N$ theory on $S^d$ admits a sequence of saddle points with constant $\sigma=-k(k+1)$ and $k$ a positive integer. 
We have identified the $k=1$ saddle point as the source of leading non-perturbative effects in the large $N$ limit of the $O(N)$ model with $4<d<6$. 
The corresponding instanton solutions are well-known in the limits where  $d$ approaches $4$ and $6$ so that the theory becomes weakly coupled \cite{McKane:1978md,Mckane:1978me,McKane:1984eq}, as reviewed in Section \ref{CLASSICAL}.
In this section, we collect for completeness a number of results about the $k>1$ solutions, including 
 their ``classical actions" and the spectrum of fluctuations around them. In particular, we present new (as far as we know) classical solutions on $S^4$ and $S^6$
 where the fields $\phi^i$ are not constant, but are rather proportional to any spherical harmonic.

\subsection{Saddle points in the large $N$ theory}
\label{k-saddle}

The value of the effective action evaluated on the large $N$ saddle point is given by
\begin{equation}
S_{\sigma=-k(k+1)} = \frac{N}{2} \sum_{n=0}^{\infty} D_{n}(d) \log \Big(n(n+d-1)+\frac{1}{4}d(d-2)-k(k+1)\Big)=N f_k(d) \,,
\end{equation}
where $f_k(d)$ may be given the integral representation in (\ref{f-alpha-1}) or (\ref{f-alpha}), for $\alpha=k$. Using the recursion relation for degeneracies 
  \begin{align}
D_{n}(d) = D_{n-1}(d)+D_{n}(d-1)\,,
\end{align}
it is not hard to show that 
\begin{align}
f_{k}(d) -f_{k-1}(d) = \sum_{m=0}^{k-1}f_{m}(d-2)\,.
\end{align}
From this relation we find 
 \begin{align}
f_{k}(d) =f_{0}(d) +\sum_{m=0}^{k-1}(k-m)f_{m}(d-2)\,,
\end{align}
and finally we obtain
 \es{SGenLargeN}{
S_{\sigma = -k (k+1)} - S_{\sigma = 0} = N \left( f_{k}(d)  - f_0(d) \right) = N\sum_{m=1}^{k} {{k+m}\choose{2m}}F_{S^{d-2m}}^\text{free scalar}\,,
 }
where we used the fact that $f_0(d) = F_{S^d}^\text{free scalar}$, with an explicit expression for general $d$ given in \eqref{Ffreesc}. For example, for the $k=2$ saddle, we find that at large $N$, the effective action evaluated on the saddle point is
\begin{align}
S_{\sigma = -6} - S_{\sigma = 0} =  \left( 3 F_{S^{d-2}}^\text{free scalar}+F_{S^{d-4}}^\text{free scalar}\right) N \,.
\end{align}

Let us now consider the spectrum of fluctuations of $\sigma$ around these saddles. For this we need to diagonalize the ``kinetic" operator $-NG_{k}^{2}(\vec{x}_{1},\vec{x}_{2})$, where $G_k(\vec{x}_1,\vec{x}_2)$ is the $\phi$'s Green's function around the $\sigma=-k(k+1)$ solution. Using Eq.~(\ref{GGen}) and hypergeometric function relations we find 
\begin{align}
G_{k}(\vec{x}_{1},\vec{x}_{2}) = \frac{\Gamma(\frac{d}{2}-1)}{4\pi^{d/2}}\frac{1}{s^{d-2}}\,_{2}F_{1}\Big(-k,k+1,2-\frac{d}{2},\frac{s^{2}(\vec{x}_{1},\vec{x}_{2})}{4}\Big)\,,
\end{align}
where we fixed the constant $C$ as before by noting that $G_{k}(\vec{x}_{1},\vec{x}_{2})$ must have the same 
short-distance singularity as the propagator of a conformally coupled scalar.   As for the $k=1$ saddle, the Green's function can be expanded in spherical harmonics $Y_{n,\vec{m}}(x)$ 
with eigenvalues $\lambda_{n}^{(k)}$
\begin{align}
-N G_{k}^{2}(\vec{x}_{1},\vec{x}_{2})= \sum_{n,\vec{m}}\lambda_{n}^{(k)} Y^{*}_{n,\vec{m}}(\vec{x}_{1})Y_{n,\vec{m}}(\vec{x}_{2})\,.
\end{align}
Using that 
\begin{align}
\frac{1}{s(\vec{x}_{1},\vec{x}_{2})^{2\Delta}}= \sum_{n=0}^{\infty}k_{n}(\Delta) Y_{n,\vec{m}}^{*}(\vec{x}_{1})Y_{n,\vec{m}}(\vec{x}_{2}), \quad k_{n}(\Delta) =
\pi^{\frac{d}{2}}2^{d-2\Delta}\frac{\Gamma(\frac{d}{2}-\Delta)\Gamma(n+\Delta)}{\Gamma(\Delta) \,,\Gamma(d+n-\Delta)}
\end{align}
we find for general $k$ that
\begin{align}
\frac{\lambda_{n}^{(k)}}{\lambda_{n}^{(0)}} =\prod _{m=1}^k \frac{(d+n+2 m-2)(n-2m+1) }{ (d+n-2 m-2)(n+2m+1)}, \quad \lambda_{n}^{(0)} = 
\frac{N}{2^{d}(4\pi)^{\frac{d-3}{2}}\sin(\frac{\pi d}{2})\Gamma(\frac{d-1}{2})}\frac{\Gamma(d+n-2)}{\Gamma(n+2)}\,.
\end{align}
Note that $\lambda_{n}^{(0)}$ is positive for all $n$, whereas 
$\lambda_{n}^{(k)}$  are negative for $n=0,2,4,\ldots, 2k-2$ and zero for $n=1,3,5,\ldots,2k-1$ with degeneracies $D_{n}(d)$. In the case of $k=1$, i.e.~the physical instanton which was the focus of the paper, we get a single negative mode and $d+1$ zero modes. For $k>1$ there are several negative modes as well as additional zero modes, making the physical interpretation of these solutions unclear.  For instance, for $k=2$, one finds $(d+1)(d+2)/2$ negative modes and $(d+1)(d+2)(d+3)/6$ zero modes. 

\subsection{Expansion around $6$ and $4$ dimensions}

Let us now discuss how these saddles behave close to $d=6$ and $d=4$.
First note that if we take $d = 6 - \epsilon$ in Eq.~\eqref{SGenLargeN}, and expand to leading order in $\epsilon$, then only a finite number of terms contribute to the sum, because the general formula \eqref{Ffreesc} for the $S^d$ free energy implies to leading order in $\epsilon$
 \es{FSdEpsilon}{
  F_{S^{6 - \epsilon - 2m}}^\text{free scalar} 
   = \begin{cases}
    \frac{1}{90 \epsilon} \,, & m=1 \\
    -\frac{1}{3 \epsilon} \,, & m=2 \\
    - \frac{2}{\epsilon} \,, & m = 3 \\
    0 \,, & m \geq 4  \,.
   \end{cases}
 } 
Thus we find that, in the large $N$ theory, the value of the action evaluated on the $\sigma = -k(k+1)$ saddle point is
 \es{Sd6}{
  S_{\sigma = -k(k+1)}  - S_{\sigma = 0}\bigg|_{d=6 - \epsilon} 
   = - N \frac{k^2 (k+1)^2 (k^2 + k - 3)}{360 \epsilon}
 }
in $d = 6 - \epsilon$ dimensions.  Performing a similar computation in $d = 4 + \epsilon$ dimensions, we find 
 \es{Sd4}{
  S_{\sigma = -k(k+1)} - S_{\sigma = 0} \bigg|_{d=4 + \epsilon} 
   = N \frac{k^2 (k+1)^2 }{12 \epsilon} \,.
 }
The fact that these expressions are proportional to $1/ \epsilon$ suggests that we should be able to find all the $k$-saddles in perturbation theory in $\epsilon$.
Indeed, in the following we find analogous classical solutions in the theories \eqref{Cubic6d} and \eqref{ON4d} in $d=6-\epsilon$ and $d=4+\epsilon$, respectively, at values of $N$ that are not necessarily large.

\subsection{Instantons close to $d=6$}

Let us start with the cubic theory \eqref{Cubic6d} in $d=6$ at arbitrary couplings $g_1$ and $g_2$, conformally mapped to $S^6$ as in \eqref{S6-action}.  Due to conformal symmetry at the classical level, any solution to the classical equations of motion for the $S^6$ theory \eqref{S6-action}, 
 \es{ClassicalS6}{
  \nabla^2 \phi^i &= (6 + g_1 \overline{\sigma}) \phi^i \,, \\
  \nabla^2 \overline{\sigma} &= \frac{g_1}{2} \phi^i\phi^i + \frac{g_2}{2} \overline{\sigma}^2 + 6 \overline{\sigma}
 }
can be mapped to a solution to the classical equations of motion \eqref{Class-EOM} of the $\R^6$ theory \eqref{Cubic6d}, and vice versa.   In \eqref{ClassicalS6}, we relabeled $\sigma \to \overline{\sigma}$ when we derived the equations of motion from \eqref{S6-action} in order not to confuse $\overline{\sigma}$, which appears in the 6d theory with the $\sigma$ field from the large $N$ theory.  The two fields differ by an overall normalization factor.  

The feature of the solutions we want to find is that $\overline{\sigma}$ is constant on $S^6$,\footnote{There should also be solutions where $\overline{\sigma}$ is not a constant.  Some of them can be obtained by performing a conformal transformation to flat space, followed by a translation and dilation, followed by a conformal transformation back to the sphere.  We leave to future work an investigation as to whether this is a full set of solutions of \eqref{ClassicalS6}.\label{NonConstantFootnote}} so we may ask whether \eqref{ClassicalS6} admit such solutions.  It is easy to see that the first equation in \eqref{ClassicalS6} then takes the form of an eigenvalue equation for the operator $\nabla^2 - 6$, with $g_1 \overline{\sigma}$ being the eigenvalue.  Apart from the trivial solutions with $\phi_i =0$, non-trivial solutions exist only when $6 + g_1 \overline{\sigma}  = - n (n+5)$ for some $n = 0, 1, 2, \ldots$.  They are the $S^6$ spherical harmonics $Y_{n, \vec{m}}$.  Thus 
 \es{GotbarSigma}{
  \overline{\sigma} = - \frac{n(n+5) + 6}{g_1} = - \frac{(n+2)(n+3)}{g_1} \,.
 }
Comparing the coefficients of $\sigma \phi^i \phi^i$ in the $S^6$ action \eqref{ClassicalS6} and the large $N$ action \eqref{LargeNaction-Sd}, we see that $\sigma = g_1 \overline{\sigma}$.  Therefore, the solution \eqref{GotbarSigma} precisely matches $\sigma = -k(k+1)$ upon the identification $k = n+2$.  The second equation in \eqref{ClassicalS6} is solved provided that $\phi^i \phi^i$ is a constant equal to\footnote{Since the right-hand side is negative, these solutions involve imaginary $\phi_i$. This is similar to the instanton solutions in the $O(N)$ model in 
$4-\epsilon$ dimensions, which are also imaginary (see \cite{McKane:1978md}, Section \ref{Inst-4d}, and the following section).} 
 \es{phiSQ}{
  \phi^i  \phi^i = - \frac{g_2 \overline{\sigma}^2 + 12 \overline{\sigma}}{g_1} \,.
 }
It is not hard to arrange for the $\phi^i$ to be proportional to the spherical harmonics with index $n$ and at the same time for $\phi^i \phi^i$ to be constant.  Indeed, if $D_n$ is the number of linearly independent such spherical harmonics, and $\{ Y_{n, p} \}$, with $p = 1, \ldots, D_n$ is an orthonormal  basis of real spherical harmonics, then take
 \es{phiiChoice}{
  \phi^i(\vec{x}) = \begin{cases}
    \alpha \sqrt{\frac{\Vol(S^6)}{D_n}} Y_{n, i}(\vec{x}) \,, &i = 1, \ldots, D_n \,, \\
    0 \,, &i = D_{n} + 1, \ldots, N \,,
  \end{cases}
 }
where $\alpha$ is a constant.  Then $\phi^i \phi^i = \alpha^2$ is indeed a constant on $S^6$.  The choice \eqref{phiiChoice} is not unique because one can perform $O(N)$ rotations on the $N$ scalars.\footnote{Because $O(N - D_n)$ transformations acting on the last $N - D_n$ scalars leave the solution \eqref{phiiChoice} invariant, the moduli of the more general solution obtained by acting with $O(N)$ rotations on \eqref{phiiChoice} parameterize the coset space $O(N) / O(N-D_n)$.  Explicitly, we can parameterize $O(N)/O(N - D_n)$ with $D_n$ orthonormal $N$-component vectors $u_p^i$, with $p = 1, \ldots, D_n$ and $i = 1, \ldots N$, obeying $u_p^i u_q^i = \delta_{pq}$.  Then, the general solution obtained by acting with an $O(N)$ rotation on \eqref{phiiChoice} can be written as $\phi^i(\vec{x}) = \alpha \sqrt{\frac{\Vol(S^6)}{D_n}} u^i_p Y_{n, p}(\vec{x})$.\label{ModuliFootnote}}  For example, when $n=0$, we have $\phi^i = \alpha (1, 0, 0, \ldots)$, up to $O(N)$ rotations.  For $n=1$, also up to $O(N)$ rotations, we have $\phi^i = \alpha (\hat x_1, \hat x_2, \ldots, \hat x_7, 0, 0, \ldots$), where $\hat x_i$ are the components of the unit vector $\hat x$ in $\R^7$ parameterizing $S^6$.  Note that for constructing such solutions, we require $N \geq D_n$.

The on-shell action of these solutions can be obtained by simply plugging \eqref{GotbarSigma} and \eqref{phiSQ} in the action \eqref{S6-action} and using $-\nabla^2 \phi^i = n (n+5) \phi^i$.  This calculation gives
 \es{S6dn}{
  S_n = \frac{8 \pi^3 (n + 2)^2 (n + 3)^2 \left[ 18 g_1 - (n + 2)(n+3) g_2 \right] }{45 g_1^3} \,.
 } 

The above analysis holds for the classical theory \eqref{S6-action} for any values of the couplings $g_1$ and $g_2$.  The same analysis can also be applied to the critical theory in $6-\epsilon$ dimensions, by simply plugging into \eqref{S6dn} the critical values of the coupling given in Eq.~\eqref{gstar}.  After doing so, one obtains
 \es{Sn6dmeps}{
  S_n = - \frac{(n+2)^2 (n+3)^2 (n^2 + 5n + 3)}{360 \epsilon} N 
   - \frac{(n+2)^2 (n+3)^2 (8 n^2 + 40 n + 59)}{30 \epsilon} + O(1/N) \,.
 }
(An exact expression in $N$ can also be obtained if one uses the exact formulas for $g_1^*$ and $g_2^*$ from \cite{Fei:2014yja}.)  We see that this expression precisely matches \eqref{Sd6} after we identify $k = n + 2$, as explained right below \eqref{GotbarSigma}.  

\subsection{Instantons close to $d=4$}

One can similarly find additional classical instantons in $d=4$ in the quartic scalar model \eqref{S4-action}.  
The solutions to the classical equations of motion for the $S^4$ and $\R^4$ theories are in one-to-one correspondence, and we choose to work on $S^4$.  Instead of using the quartic action \eqref{S4-action}, it is convenient to write the theory on $S^4$ with the help of the auxiliary Hubbard-Stratonovich field $\sigma$:
\begin{equation}
S = \int d^4x\sqrt{g}\left(\frac{1}{2}\partial^{\mu}\phi^i\partial_{\mu}\phi^i 
+ \frac 12 \phi^i\phi^i \left (2 + \sigma \right )
- \frac{\sigma^2}{4g} \right)\,.
\label{S4-actionnew}
\end{equation}
Performing the path integral over $\sigma$ (or, at the classical level, solving for $\sigma$ from its equation of motion and plugging the solution back into \eqref{S4-actionnew}), one recovers the quartic model in \eqref{S4-action}, so the theories \eqref{S4-action} and \eqref{S4-actionnew} are indeed equivalent.  The equations of motion following from \eqref{S4-actionnew} are
 \es{eomsS4}{ 
   \sigma &= g \phi^i \phi^i \,, \\
  \nabla^2 \phi^i &= (2 + \sigma)  \phi^i \,.
 }

As in 6-d, we should look for solutions for which $\sigma$ is a constant on $S^4$.\footnote{Same comment as in Footnote~\ref{NonConstantFootnote}.}  The equation \eqref{eomsS4} then takes the form of an eigenvalue problem for the operator $\nabla^2 - 2$, with $\sigma$ being the eigenvalue.  The solutions are the $S^4$ spherical harmonics, which have eigenvalues of the Laplacian equal to $-n(n+3)$, with $n = 0, 1, 2, \ldots$.  This implies
 \es{phiSQS4}{
  \sigma = - n(n+3) - 2 = - (n+1)(n+2) \,.
 }
This expression precisely matches the solutions $\sigma = -k(k+1)$ in the large $N$ theory upon the identification $k = n+1$.  As in 6-d, we can write explicit solutions for $\phi^i$ obeying both the constraint $\phi^i \phi^i = 1/g$ and solving \eqref{eomsS4}. If, for a given $n$, we consider a real orthonormal basis $\{ Y_{n, p}\} $, $p = 1, \ldots, D_n$, for the spherical harmonics with Laplacian eigenvalue $-n(n+3)$ whose degeneracy is $D_n$, then we can take
  \es{phiiChoiceS4}{
  \phi^i(\vec{x}) = \begin{cases}
     \sqrt{\frac{\sigma \Vol(S^4)}{g D_n}} Y_{n, i}(\vec{x}) \,, &i = 1, \ldots, D_n \,, \\
    0 \,, &i = D_{n} + 1, \ldots, N \,,
  \end{cases}
 }
up to $O(N)$ rotations.\footnote{As in Footnote~\ref{ModuliFootnote}, we can write the general solution obtained after acting with $O(N)$ transformations on \eqref{phiiChoiceS4} using a set of $D_n$ orthonormal $N$-component vectors $u^i_p$, with $p = 1, \ldots, D_n$ and $i = 1, \ldots, N$.  The general solution is $\phi^i(\vec{x}) = \sqrt{\frac{\sigma \Vol(S^4)}{g D_n}} u^i_p Y_{n, p}(\vec{x})$.}  In fact, the solution with $n=0$ is precisely the solution \eqref{inst-4d-const} considered in Section~\ref{Inst-4d}.

The classical action evaluated on the instanton solution presented above can be obtained by simply plugging \eqref{phiSQS4} into the action \eqref{S4-actionnew} and noticing that after using the $\phi^i$ equation of motion, only the last term in \eqref{S4-actionnew} survives.  Thus, one obtains
 \es{Snd4}{
  S_n = - \frac{2 \pi^2 (n+1)^2 (n+2)^2}{3 g} \,.
 }

The analysis above was in the classical theory \eqref{S4-actionnew} (or equivalently \eqref{S4-action}) for any values of $g$.  To study the critical theory in $4+\epsilon$ dimensions, we can simply plug the value of the critical coupling \eqref{gstar4} into \eqref{Snd4}, obtaining
 \es{Sn4dpeps}{
  S_n =  \frac{ (n+1)^2 (n+2)^2(N+8)}{12\epsilon} \,.
 }
At leading order in large $N$, this expression matches \eqref{Sd4}, provided that we identify $k = n+1$, as required to match \eqref{phiSQS4} with $\sigma = -k(k+1)$.

\section*{Acknowledgments}

We are very grateful to Lin Fei for an extensive collaboration at the early stages of this project. 
We also thank L.~Iliesiu, I.~Papadimitriou, G.~Parisi, A.~Petkou, V.~Rychkov, and E.~Witten for useful discussions.
This research was supported in part by the US NSF under Grants No.~PHY-1620059 (IRK), PHY-1620542 (SG), PHY-1820651 (SSP), and PHY-1914860 (SG and IRK), and in part by the Simons Foundation Grant No.~488653 (SSP).  SSP was also supported in part by an Alfred P.~Sloan Research Fellowship.  
GT was supported by  the MURI grant W911NF-14-1-0003 from ARO, by DOE grant DE-SC0007870 and by DOE Grant No. DE-SC0019030.  IRK is also grateful to the Mainz Institute for Theoretical Physics for the hospitality and opportunity to lecture on this subject at the 2019 MITP Summer School.

\appendix

\section{Computation on $S^{d-1} \times \R$}

On $S^4 \times \R$ we write the partition function as
 \es{S4RZ}{
  Z_{S^{d-1} \times \R}
   = \tilde A_0(d) e^{- N \tilde f_0(d)} \left[ 1 + \frac{\tilde A_1(d)}{\tilde A_0(d)} e^{- N (\tilde f_1(d) - \tilde f_0(d))} + \cdots \right] \,,
 }
where we put tildes on all quantities so we do not confuse them with the analogous quantities in the $S^d$ computation.

To place the theory on $S^{d-1} \times \R$ we proceed as follows.  We use coordinates $(t, \hat p)$ for parameterizing this space, and $(\tau, \rho, \hat n)$ for parameterizing the moduli space.  Then take:
 \es{PXS4R}{
  P &= \begin{pmatrix} \sinh t \ , \ \hat p \ , \ \cosh t \end{pmatrix} \,, \\
  X &= \begin{pmatrix}  \cosh \rho \sinh \tau \ , \  \hat n \sinh \rho \ , \ \cosh \rho  \cosh \tau \end{pmatrix}  \,.
 }
In these coordinates, the instanton profile is
 \es{InstProfileS4R}{
  \sigma = - \frac{2}{(\cosh \rho \cosh (\tau  - t)  - \sinh \rho\, \hat p \cdot \hat n )^2}
 } 
The most symmetric configuration is at $\rho = \tau = 0$, where 
 \es{SigmaSymmetric}{
  \sigma = - \frac{2}{\cosh^2 t } \,.
 }
We will do the computations at this point in the instanton moduli space.

\subsection{$\phi$ determinant}

The operator whose eigenvalues we should compute is
 \es{OpS4R}{
  - \partial_t^2 - \frac{2}{\cosh^2 t} - \nabla_{S^{d-1}}^2 + \frac{(d-2)^2}{4}  \,.
 }
To evaluate it, we can use the formula that the regularized determinant of the operator
 \es{OpSech}{
  {\cal O}(k) = - \partial_t^2 - \frac{k(k+1)}{\cosh^2 t} + a^2 \,,
 }
which is
 \es{det}{
  \det {\cal O}(k) =  \frac{\Gamma(a + 1) \Gamma(a)}{\Gamma(a + k + 1) \Gamma( a- k)} \,.
 } 
For the case of interest to us, $k=1$, this gives
 \es{det1}{
  \det {\cal O}(1) = \frac{a-1}{a + 1} \,.
 }

For us, we have modes with $a^2 = n(n+d - 2) +  \frac{(d-2)^2}{4}  = \left( n + \frac{d-2}{2} \right)^2$, so $a = n + \frac{d-2}{2}$ with degeneracy $D_{n-1}(d-1)$. Then we find
 \es{efS4R}{
  \tilde f_1(d) - \tilde f_0(d) = \frac 12 \sum_{n=0}^\infty D_n(d-1) \log \frac{ n + \frac{d-2}{2} - 1  }{ n + \frac{d-2}{2} + 1 } \,.
 }
This can be rewritten as
 \es{efS4RAgain}{
   \tilde f_1(d) - \tilde f_0(d) = \frac 12 \sum_{n=0}^\infty \biggl[ D_n(d-1) - D_{n-2}(d-1) \biggr]  \log \left( n + \frac{d-2}{2} - 1 \right)  \,.
 } 
Quite nicely, $D_n(d-1) - D_{n-2}(d-1) = D_n(d-2) + D_{n-1}(d-2)$ so this expression is identical to \eqref{f1mf0Again}.  Thus
 \es{f1f0Final}{
 \tilde f_1(d) - \tilde f_0(d)  = F_{S^{d-2}}^{\text{free scalar}} \,.
}

\section{Computation of one-loop determinants}
\label{oneloopdet}
In order to compute various  one-loop determinants it is useful to define a  general function $f_{\alpha}(d)$ as 
\begin{align}
f_{\alpha}(d)= \frac{1}{2}\sum_{n=0}^{\infty}D_{n}(d )\log\Big[\big(n+\frac{d}{2}+\alpha\big)\big(n+\frac{d}{2}-\alpha-1\big)\Big]\,, \label{deff1}
\end{align}
where the degeneracies $D_{n}(d)$ are given by $D_{n}(d)=\frac{(2n+d-1)\Gamma(n+d-1)}{n! \Gamma(d)}$. For integer $\alpha=k$ the function $f_{\alpha}(d)$
coincides with the ``classical action" of the $\sigma_{k}=-k(k+1)$ saddle point in the large $N$ theory. For $k=1$ we have the instanton discussed in the main text of the paper, and $k=0$ just corresponds to the perturbative vacuum, so that $f_{\alpha=0}$ is free energy of a free conformal scalar.  To compute this function, we can differentiate it by $\alpha$ to obtain 
\begin{align}
\frac{\partial f_{\alpha}(d)}{\partial \alpha}= -\frac{(\alpha+\frac{1}{2})}{\Gamma(d)}\sum_{n=0}^{\infty} \frac{\Gamma(n+d-1)}{n!}\Big(\frac{1}{n+\frac{d}{2}+\alpha}+\frac{1}{n+\frac{d}{2}-\alpha-1}\Big)\,.
\end{align}
Then using the formula 
\begin{align}
\sum_{n=0}^{\infty} \frac{\Gamma(n+a)}{n!} \frac{1}{n+b} = \frac{\pi}{\sin \pi a}\frac{\Gamma(b)}{\Gamma(1-a+b)} \label{sum1}
\end{align}
we finally get 
\begin{align}
\frac{\partial f_{\alpha}(d)}{\partial \alpha}=  \frac{\pi (\alpha+\frac{1}{2})}{\Gamma(d)\sin \pi d} \left(\frac{\Gamma(\frac{d}{2}+\alpha)}{\Gamma(\alpha+2-\frac{d}{2})}+\frac{\Gamma(\frac{d}{2}-\alpha-1)}{\Gamma(1-\frac{d}{2}-\alpha)}\right)\,.
\end{align}
Therefore integrating over $\alpha$ we find 
\begin{align}
f_{\alpha}(d)=f_{0}(d)+ \int_{0}^{\alpha} dx \frac{\pi (x+\frac{1}{2})}{\Gamma(d)\sin \pi d} \left(\frac{\Gamma(\frac{d}{2}+x)}{\Gamma(x+2-\frac{d}{2})}+\frac{\Gamma(\frac{d}{2}-x-1)}{\Gamma(1-\frac{d}{2}-x)}\right)\,.
\label{f-alpha-1}
\end{align}
After using the identity $\Gamma(x)\Gamma(1-x)=\pi/\sin(\pi x)$, this may be also rewritten as
\begin{equation}
\begin{aligned}
&f_{\alpha}(d)=f_{0}(d)+\frac{1}{\sin(\frac{\pi d}{2})\Gamma\left(d-1\right)}\int_{0}^{\alpha} dx\,x \sin(\pi x)  \Gamma\left(\frac{d-2}{2}+x\right)\Gamma\left(\frac{d-2}{2}-x\right)\\
&~~~~~~~~~~~~~~~~~
-\frac{1}{\sin(\frac{\pi d}{2})\Gamma\left(d\right)} \int_{-1/2}^{\alpha-1/2} dx\,x \cos(\pi x) \Gamma\left(\frac{d-1}{2}+x\right)\Gamma\left(\frac{d-1}{2}-x\right)
\,.
\label{f-alpha}
\end{aligned}
\end{equation}
Applying this formula to the case $\alpha=1$, corresponding to the $\sigma=-2$ instanton, we see that the second integral in the formula above vanishes, and the first integral reproduces the free energy of a free conformal scalar (\ref{Ffreesc}) on $S^{d-2}$, so that 
\begin{equation}
f_{1}(d)=f_{0}(d)+f_{0}(d-2)\,,
\end{equation}
in agreement with (\ref{f1mf0Final}). 

It is also useful to define another function 
\begin{align}
\tilde{f}_{\alpha}(d)=  \frac{1}{2}\sum_{n=2}^{\infty}D_{n}(d )\log\Big[\big(n+\frac{d}{2}+\alpha\big)\big(n+\frac{d}{2}-\alpha-1\big)\Big]\,, \label{defft}
\end{align}
where the sum goes over $n$ from $2$ to $+\infty$ (i.e., in the case of the instanton solution, it excludes the negative and zero modes).  The computation of $\tilde{f}_{\alpha}(d)$ is similar to the formulas above, we just have to exclud  in (\ref{sum1}) the $n=0$ and $n=1$ terms. We obtain 
\begin{align}
\tilde{f}_{\alpha}(d)=&\tilde{f}_{0}(d)+ \int_{0}^{\alpha} dx(x+\frac{1}{2}) \bigg( \frac{\pi }{\Gamma(d)\sin \pi d}\bigg(\frac{\Gamma(\frac{d}{2}+x)}{\Gamma(x+2-\frac{d}{2})}+\frac{\Gamma(\frac{d}{2}-x-1)}{\Gamma(1-\frac{d}{2}-x)}\bigg)\notag\\
&\quad +\frac{d+1}{(\frac{d}{2}-x ) (\frac{d}{2}+x +1)}+\frac{1}{ (\frac{d}{2}+x)( \frac{d}{2}-x-1)}\bigg)\,. \label{ftres}
\end{align}
The integral over $x$ may have some divergencies, which  should be regularized by taking principal value of the integral. Note that the ratio of determinants $\log R_{d}$, defined in Section \ref{sigmadet}, can be written as  
\begin{align}
\log R_{d} = \tilde{f}_{\frac{d}{2}-4}(d)- \tilde{f}_{d/2}(d) \,.
\end{align}
So, using the integral (\ref{ftres}) for $d=5$ one finds 
\begin{align}
\log R_{d=5} = -\frac{\zeta(3)}{\pi^{2}} +\log \frac{9 \sqrt{15}}{2^{9}\pi^{3}}\,.
\end{align}
which is the result given in Section \ref{sigmadet}.

\subsection{Fluctuation determinants in $d=6-\epsilon$}
\label{det-6d}
We can now apply the above formulas to the calculation of determinants of fluctuations around the classical solution (\ref{sigmac}) in the cubic theory in $d=6-\epsilon$. The analogous calculation for the case of a single scalar field with cubic potential was carried out in \cite{Mckane:1978me}. To support the interpretation of the solution (\ref{sigmac}) as the instanton responsible for tunneling from the metastable ground state, it is important to check that even in the presence of the additional $N$ fields $\phi^i$, there is still a single negative mode (as well as $d+1$ zero modes) as we now show. 

Following the discussion in Section \ref{Inst-6d}, we know that for the cubic theory, the instanton solution is constant when mapped to $S^d$. Thus, the fluctuation around the classical solutions requires calculating the determinants of the following operators:
\begin{align}
&M_{\sigma} = -\nabla^2_{S^d}+\frac{d(d-2)}{4}-g_2 \sigma_c =  -\nabla^2_{S^d}+\frac{d(d-2)}{4}-12 \,, \notag\\
&M_{\phi} = -\nabla^2_{S^d}+\frac{d(d-2)}{4}-g_1 \sigma_c =-\nabla^2_{S^d}+\frac{d(d-2)}{4}-12 z\,,
\end{align}
where $\sigma_c = -12/g_{2}$, as discussed in the Section \ref{Inst-6d}, and $z=g_{1}/g_{2}$. 

Let us first examine the presence of negative eigenvalues. For the $\sigma$ fluctuations, the calculation is identical to the one in \cite{Mckane:1978me}, and for $\sigma=6-\epsilon$, we find a single negative mode for $n=0$. Let us check that there are no additional negative modes coming from the $\phi$ fluctuations. The eigenvalues of $M_{\phi}$ at $d=6-\eps$ are:
\begin{equation}
\lambda_n = \frac{(n+5/2-\eps/2+\zeta)(n+5/2-\eps/2-\zeta)}{(n+2-\eps/2)(n+3-\eps/2)}, \quad n=0,1,2,\ldots
\end{equation}
where we defined $\zeta =\frac{1}{2}\sqrt{1+48 z}$. We see that in order for all eigenvalues to be positive, we must require:
\begin{equation}
\zeta< 5/2 - \eps/2 \,.
\end{equation}
Or, equivalently,
\begin{equation}
z=\frac{g_1}{g_2}< 1/2 + \mathcal{O}(\eps)
\end{equation}
However, the analysis of perturbative fixed points in the cubic $O(N)$ theory \cite{Fei:2014yja, Fei:2014xta} shows that the ratio $g_1/g_2$ varies from $1/6$ at $N=\infty$ to about $1/8.9$ at $N=N_{crit}$. Therefore, $M_{\phi}$ does not contribute additional negative modes. 

The determinants of the fluctuation operators may be evaluated explictly using the functions defined in the previous section. Excluding the $n=0$ and $n=1$ modes which may be treated separately, we can compute the ratio of determinants:
\begin{align}
&\frac{1}{2} \log\left(\frac{ \det' M_{\sigma}^{(1)}}{ \det' M_{\sigma}^{(0)}}\right) = \frac{1}{2} \sum_{n=2}^{\infty} D_{n}(d) \log \frac{(n+\frac{d}{2}-4)(n+\frac{d}{2}+3)}{(n+\frac{d}{2}-1)(n+\frac{d}{2})} = \tilde{f}_{3}(d)-\tilde{f}_{0}(d) \,, \notag\\
&\frac{1}{2} \log \left(\frac{ \det' M_{\phi}^{(1)}}{ \det' M_{\phi}^{(0)}} \right)= \frac{1}{2} \sum_{n=2}^{\infty} D_{n}(d) \log \frac{(n+\frac{d}{2}-\frac{1}{2}+\zeta)(n+\frac{d}{2}-\frac{1}{2}-\zeta)}{(n+\frac{d}{2}-1)(n+\frac{d}{2})}= \tilde{f}_{\zeta-1/2}(d)-\tilde{f}_{0}(d) \,,
\end{align}
where the superscripts `$(1)$' and `$(0)$' refer to the instanton and perturbative saddles, respectively, and $\det'$ means we are excluding $n=0,1$. We can first calculate $\log \left(\frac{ \det' M_{\phi}^{(1)}}{ \det' M_{\phi}^{(0)}}\right) $, and then setting $\zeta=7/2$ will give us the corresponding result for the $\sigma$ determinant.
The explicit computation using the integral in eq. (\ref{ftres}) gives at $d=6-\epsilon$:
\begin{align}
 \log\left(\frac{ \det' M_{\phi}^{(1)}}{ \det' M_{\phi}^{(0)}}\right)  =&-\frac{\left(1-4 \zeta ^2\right)^2 \left(4 \zeta ^2-13\right)}{23040 \epsilon }-\frac{1}{1382400}(2 \zeta -1) \big(4384 \zeta ^5-2416 \zeta ^4-23408 \zeta ^3\notag\\
&-32824 \zeta ^2-138614 \zeta -898747\big)\notag\\
&-\frac{1}{2} \log (2 \zeta +5)-\frac{7}{2} \log (2 \zeta +7)+11 \log 2+\frac{1}{2}\log 3 \notag\\
&+\frac{1}{240}\int_{0}^{\zeta-\frac{1}{2}}d\alpha (\alpha-1)\alpha(\alpha+1)(\alpha+2)(2\alpha+1)(H_{3-\alpha}+H_{\alpha})\,,
\end{align}
where $H_{\alpha}$ is the Harmonic number. It is possible to take explicitly the last integral, but expression is very cumbersome, so we keep it as an integral. Setting $\zeta=7/2$, we find (in agreement with \cite{Mckane:1978me}):
\begin{align}
 \log\left(\frac{ \det' M_{\sigma}^{(1)}}{ \det' M_{\sigma}^{(0)}}\right) &=-\frac{18}{5 \epsilon }-2 \zeta '(-3)-11 \zeta '(-1)-\frac{109}{240}+\frac{18 \gamma }{5}+\frac{7}{2} \log \frac{12}{7} \notag\\
&\approx -\frac{18}{5 \epsilon }+5.3192\,. \label{logMsigans}
 \end{align}
Here, the $1/\epsilon$ pole is a UV divergence that comes from the summation over large $n$ (in particular, the $n=0,1$ modes clearly do not affect this UV pole). We can also extract analytically the coefficient of the $1/\epsilon$ pole in the $\phi$ determinant. We find, in terms of $z=g_{1}/g_{2}$:
\begin{align}
\frac{1}{2} \log \left(\frac{ \det M_{\phi}^{(1)}}{ \det M_{\phi}^{(0)}}\right) |_{\textrm{pole}} =\frac{6}{5 \epsilon }(z^{2}-4 z^{3})\,.
\end{align} 
Since the theory is renormalizable, we expect this divergence to be cancelled by the perturbative renormalization of the coupling constants. Let us check this explicitly as a further test of our results. The classical action on the instanton background is 
\begin{equation}
S_{\rm class} = \frac{768\pi^3}{5 g_2^2}
\end{equation}
where $g_2$ should be viewed as the bare coupling. The renormalized coupling is related to the beta function by
\begin{equation}
\frac{\mu^{\eps}}{(g_2^2)_{\rm bare}} =\frac{1}{g_{2}^2}-\frac{2 \beta_{2,6d}(g_1,g_2)}{g_2^3}\frac{1}{\eps}
\end{equation}
where $\mu$ is a renormalization scale, and $\beta_{2,6d}$ is the beta function for $g_2$ in $d=6$. 
Using the explicit result for the perturbative beta function found in \cite{Fei:2014yja}, we get 
\begin{equation}
\begin{aligned}
\frac{\mu^{\eps}}{(g_2^2)_{\rm bare}}
=&\frac{1}{g_{2}^2} + \frac{1}{2(4\pi)^3}\left(\frac{3}{\eps}+\frac{N}{\eps}\left(4\frac{g_1^3}{g_2^3}-\frac{g_1^2}{g_2^2} \right) \right) \\
=&\frac{1}{g_{2}^2}+\frac{1}{2(4\pi)^3}\left(\frac{3}{\eps}+\frac{N}{\eps}\left(4z^3-z^2 \right) \right)  \,.
\end{aligned}
\end{equation}
We can then see that the $1/\epsilon$ pole precisely cancels the one coming from the fluctuation determinants 
\begin{align}
&\frac{1}{2}\log \left(\frac{ \det M_{\sigma}^{(1)}}{ \det M_{\sigma}^{(0)}}\right)\Big{|}_{\rm pole} +\frac{N}{2} \log \left(\frac{ \det M_{\phi}^{(1)}}{ \det M_{\phi}^{(0)}}\right)\Big{|}_{\rm pole}= -\frac{18}{5\eps} + N\frac{6}{5\eps} \left(z^2-4z^3 \right) \,,
\end{align}
as expected from renormalizability.

\section{Thermal mass on $S^1\times \mathbb{R}^{d-1}$}
\label{thermal-App}

In this section we briefly discuss the calculation of the free energy for the large $N$ critical theory on $S^1\times \mathbb{R}^{d-1}$, where $S^1$ is the thermal circle of circumference $\beta=1/T$. We will use a formal dimensional regularization so that all power-like divergences are automatically subtracted away. Starting from the Lagrangian of the critical theory (\ref{LargeNaction}) and integrating out the $\phi^i$ fields, we get a path-integral over $\sigma$ with action
\begin{equation}
S_{\sigma}=\frac{N}{2}\log{\rm det}(-\partial^2+\sigma) = N {\cal F}(\sigma)\,.
\end{equation} 
At large $N$, we can evaluate the free energy by extremizing with respect to $\sigma$, assuming the saddle point occurs for constant $\sigma$. Evaluating the $\phi$ one-loop determinant one finds
\begin{equation}
\begin{aligned}
{\cal F}(\sigma) &= \frac{1}{2}V_{d-1} \sum_{n=-\infty}^{\infty} \int \frac{d^{d-1}p}{(2\pi)^{d-1}} \log \left((\frac{2\pi n}{\beta})^2+p^2+\sigma\right) \,,
\label{F-sigma}
\end{aligned}
\end{equation}
where $V_{d-1}$ is the (infinite) volume of the plane $\R^{d-1}$. To recover the result for the free theory, one should set $\sigma=0$ in this expression, which gives, using dimensional regularization throughout
\begin{equation}
\begin{aligned}
{\cal F}_{\rm free} &=\frac{1}{2}V_{d-1} \sum_{n=-\infty}^{\infty} \int \frac{d^{d-1}p}{(2\pi)^{d-1}} \log \left((\frac{2\pi n}{\beta})^2+p^2\right)=-\frac{1}{2}V_{d-1} \sum_{n=-\infty}^{\infty} \int_0^{\infty} \frac{dt}{t} \int \frac{d^{d-1}p}{(2\pi)^{d-1}} e^{-t \left((\frac{2\pi n}{\beta})^2+p^2\right)}\\
&= -\frac{1}{2}V_{d-1} T^{d-1}\frac{1}{(4\pi)^{\frac{d-1}{2}}} \Gamma\left(\frac{1-d}{2}\right)\sum_{n=-\infty}^{\infty}\left[4\pi^2 n^2\right]^{\frac{d-1}{2}}=-\pi^{\frac{d-1}{2}} V_{d-1}T^{d-1}\Gamma\left(\frac{1-d}{2}\right)\zeta(1-d)\\
&=-V_{d-1} T^{d-1} \frac{\Gamma\left(\frac{d}{2}\right) \zeta(d)}{\pi^{\frac{d}{2}}} \,,
\end{aligned} 
\end{equation}
which is the well-known result. To obtain the free energy for the interacting fixed point, we need to extremize (\ref{F-sigma}) with respect to $\sigma$. We can compute the $\sigma$ derivative as
\begin{equation}
\begin{aligned}
\frac{d{\cal F}}{d\sigma} &= \frac{1}{2}V_{d-1}\sum_{n=-\infty}^{\infty} \int \frac{d^{d-1}p}{(2\pi)^{d-1}} \frac{1}{(\frac{2\pi n}{\beta})^2+p^2+\sigma}=\frac{1}{4}V_{d-1}\beta  \int \frac{d^{d-1}p}{(2\pi)^{d-1}} \frac{\coth \frac{\beta}{2}\sqrt{p^2+\sigma}}{\sqrt{p^2+\sigma}} \,.
\end{aligned}
\end{equation}
Writing $\coth \frac{x}{2} = 1+\frac{2}{e^x-1}$, we have
\begin{equation}
\begin{aligned}
\frac{d{\cal F}}{d\sigma} &=\frac{1}{4}V_{d-1}\beta  \left[\int \frac{d^{d-1}p}{(2\pi)^{d-1}} \frac{2}{(e^{\beta\sqrt{p^2+\sigma}}-1)\sqrt{p^2+\sigma}}+\int \frac{d^{d-1}p}{(2\pi)^{d-1}} \frac{1}{\sqrt{p^2+\sigma}}\right] \,.
\end{aligned}
\end{equation}
The first integral is convergent for any $d$, and the second one can be computed by dimensional regularization. After a change of variables in the first integral, we obtain
\begin{equation}
\begin{aligned}
\frac{d{\cal F}}{d\sigma} &=\frac{1}{4}V_{d-1}\beta \left[\frac{4\beta^{2-d}}{(4\pi)^{\frac{d-1}{2}}\Gamma\left(\frac{d-1}{2}\right)}\int_{\beta\sqrt{\sigma}}^{\infty}dy \frac{\left(y^2-\beta^2\sigma\right)^{\frac{d-3}{2}}}{e^{y}-1}-\frac{d}{(4\pi)^{\frac{d}{2}}}\sigma^{\frac{d}{2}-1}\Gamma(-\frac{d}{2})\right] \,.
\label{dF-gen-d}
\end{aligned}
\end{equation}
Let us check this result in the case $d=3$. Using the integral
\begin{equation}
\int_{\beta \sqrt{\sigma}}^{\infty} \frac{dy}{e^y-1} = \frac{\beta\sqrt{\sigma}}{2}-\log \left(2\sinh\frac{\beta\sqrt{\sigma}}{2}\right) 
\end{equation}
we get
\begin{equation}
\begin{aligned}
\frac{d{\cal F}}{d\sigma} &= -\frac{V_2}{4\pi}\log \left(2\sinh\frac{\beta\sqrt{\sigma}}{2}\right)\,.
\label{F-crit-3d}
\end{aligned}
\end{equation}
So we find that the value of $\sigma$ extremizing ${\cal F}$ is 
\begin{equation}
\sqrt{\sigma^{*}} = \frac{2}{\beta}\log\left(\frac{1+\sqrt{5}}{2}\right)
\end{equation}
and, integrating (\ref{F-crit-3d}) in $\sigma$, we obtain the free energy of the 3d critical theory
\begin{equation}
\begin{aligned}
{\cal F}_{\rm crit}& = {\cal F}_{\rm free}-\frac{V_2}{4\pi} \int_0^{\sigma^*} d\sigma \log \left(2\sinh\frac{\beta\sqrt{\sigma}}{2}\right)\\
& =-V_2 T^2 \frac{\zeta(3)}{2\pi} +
V_2 T^2 \frac{1}{5}\frac{\zeta(3)}{2\pi} = -V_2 T^2 \frac{4}{5}\frac{\zeta(3)}{2\pi}
=\frac{4}{5} {\cal F}_{\rm free} \,,
\end{aligned}
\end{equation}
which is in agreement with the result of \cite{Sachdev:1993pr}. 

Let us now consider the critical theory in $d=5$. Starting from (\ref{dF-gen-d}), we need to evaluate the integral
\begin{equation}
\int_{\beta \sqrt{\sigma}}^{\infty} dy\frac{y^2-\beta^2\sigma}{e^y-1} = 2 {\rm Li}_3(e^{-\beta\sqrt{\sigma}})+2\beta\sqrt{\sigma}  {\rm Li}_2(e^{-\beta\sqrt{\sigma}}) \,.
\end{equation}
So we get
\begin{equation}
\frac{d{\cal F}}{d\sigma} = \frac{V_4}{8\pi^2\beta^2}\left[{\rm Li}_3(e^{-\beta\sqrt{\sigma}})+\beta\sqrt{\sigma}{\rm Li}_2(e^{-\beta\sqrt{\sigma}})+\frac{\beta^3\sigma^{\frac{3}{2}}}{6}\right]\,.
\label{dFds5}
\end{equation}
Hence the saddle points $\sigma^*$ are the solutions of the equation 
\begin{equation}
\label{s-saddle}
{\rm Li}_3(e^{-x})+x{\rm Li}_2(e^{-x})+\frac{x^3}{6}=0\,,\qquad x=\beta\sqrt{\sigma^*}\,,
\end{equation}
in agreement with the result of \cite{Petkou:2018ynm} obtained by different methods. 
This equation does not have any real or purely imaginary solutions, but one finds the pair of complex conjugate solutions
\begin{equation}
x^*_{\pm} = 1.17431 \pm 1.19808 i \,.
\end{equation}
If we assume that the integration contour can be taken so that it passes through both saddle points, then integrating (\ref{dFds5}) in $\sigma$ one may get a real free energy. It would be interesting to study the thermal theory further, and clarify the relation to the non-perturbative instability on $S^d$ or $\mathbb{R}\times S^{d-1}$ that we discussed in this paper.

\bibliographystyle{ssg}
\bibliography{Instanton-Bib}

\end{document}